\documentclass[preprint]{aastex631}

\newcommand{\fwidth}{8.8cm}   
\usepackage{color}
\usepackage{natbib}
\usepackage{amsmath}
\usepackage{nicefrac}
\tightenlines

\newcommand \Angstrom   {\,{\rm \AA}}

\newcommand \beq        {\begin{equation}}
\newcommand \beqa	{\begin{eqnarray}}
\newcommand \Ca         {{\rm C}}

\newcommand \aliph      {{\rm aliph.}}
\newcommand \arom       {{\rm arom.}}
\newcommand \aliphCD    {{\rm aliph.CD}}
\newcommand \aromCD     {{\rm arom.CD}}
\newcommand \CD         {{\rm CD}}
\newcommand \aromCH     {{\rm arom.CH}}
\newcommand \aliphCH    {{\rm aliph.CH}}
\newcommand \nonaromCH  {{\rm nonarom.CH}}
\newcommand \CH         {{\rm CH}}
\newcommand \CO         {{\rm CO}}
\newcommand \cm         {\,{\rm cm}}
\newcommand \cont       {{\rm cont}}
\newcommand \Da         {{\rm D}}

\newcommand \dust       {{\rm dust}}
\newcommand \eeq	{\end{equation}}
\newcommand \eeqa	{\end{eqnarray}}

\newcommand \EM         {{\rm EM}}

\newcommand \erg	{\,{\rm erg}}

\newcommand \eV 	{\,{\rm eV}}

\newcommand \FWHM       {{\rm FWHM}}
\newcommand \gas        {{\rm gas}}

\newcommand \gtsim	{\gtrsim}		 

\newcommand \Ha 	{{\rm H}}
\newcommand \He	        {{\rm He}}
\newcommand \HH	        {{\rm H}_2}

\newcommand \K  	{\,{\rm K}}
\newcommand \kB         {k_{\rm B}}
\newcommand \kms	{\,{\rm km~s}^{-1}}
\newcommand \kpc	{\,{\rm kpc}}

\newcommand \ltsim	{\lesssim}		 

\newcommand \MJy        {\,{\rm MJy}}
\newcommand \Mpc	{\,{\rm Mpc}}

\newcommand \NH         {N_{\rm H}}

\newcommand \ppm        {\,{\rm ppm}}

\newcommand \s	        {\,{\rm s}}

\newcommand \sr  	{\,{\rm sr}}

\newcommand{\btdnote}[1]{}

\newcommand{\omittext}[1]{}
\newcommand{\newtext}[1]{{\color{blue}#1}}

\countdef\decade=200
\advance\decade by \year
\countdef\hours=201
\advance\hours by \time
\divide\hours by 60
\countdef\mins=202
\advance\mins by \hours
\multiply\mins by 60
\multiply\hours by 100
\countdef\miltime=203
\advance\miltime by \hours
\advance\miltime by \time
\advance\miltime by -\mins

\begin{document}

\title{%
        \vspace*{-2.0em}
        {\normalsize\rm Published in {\it The Astrophysical Journal Letters},
        {\bf984}:L42 (2025 May 10)}\\ 
        \vspace*{1.0em}
        {\bf Detection of Deuterated Hydrocarbon Nanoparticles in the
          Whirlpool Galaxy, M51}
	}

\author[0000-0002-0846-936X]{B.~T.~Draine}
\affiliation{Dept.\ of Astrophysical Sciences,
  Princeton University, Princeton, NJ 08544, USA}

\author[0000-0002-4378-8534]{Karin Sandstrom}
\affiliation{Department of Astronomy and Astrophysics, University of
  California San Diego, 9500 Gilman Drive, La Jolla, CA 92093, USA}

\author[0000-0002-5782-9093]{Daniel A.~Dale}
\affiliation{Department of Physics and Astronomy, University of Wyoming,
Laramie, WY 82071, USA}

\author[0000-0003-1545-5078]{J.-D.~T.~Smith}
\affiliation{Dept.\ of Physics and Astronomy, University of Toledo,
  Toledo, OH 43606, USA}


\author[0000-0001-8241-7704]{Ryan Chown}
\affiliation{Department of Astronomy, The Ohio State University, 140
  West 18th Avenue, Columbus, OH 43210, USA}

\author[0009-0001-6065-0414]{Grant P. Donnelly}
\affiliation{Ritter Astrophysical Research Center, University of
  Toledo, Toledo, OH 43606, USA}

\author[0000-0003-0014-0508]{Sara E.~Duval}
\affiliation{Dept.\ of Physics and Astronomy, University of Toledo,
  Toledo, OH 43606, USA}

\author[0000-0003-2093-4452]{Cory M.~Whitcomb}
\affiliation{Dept.\ of Physics and Astronomy, University of Toledo,
  Toledo, OH 43606, USA}


\author[0000-0002-8192-8091]{Angela Adamo} 
\affiliation{The Oskar Klein Centre, Department of Astronomy,
  Stockholm University, AlbaNova, SE-106 91 Stockholm, Sweden}

\author[0000-0003-3498-2973]{L.~Armus}
\affiliation{IPAC, California Institute of Technology, 1200
  E. California Blvd., Pasadena, CA 91125, USA}

\author[0000-0002-4153-053X]{Danielle A.~Berg}
\affiliation{Department of Astronomy, The University of Texas at
  Austin, 2515 Speedway, Stop C1400, Austin, TX 78712, USA}

\author[0000-0002-5666-7782]{Torsten B\"oker}
\affiliation{European Space Agency, Space Telescope Science Institute,
  Baltimore, MD, USA}

\author[0000-0002-5480-5686]{Alberto D.~Bolatto}
\affiliation{Department of Astronomy, University of Maryland, College
  Park, MD 20742, USA} 
\affiliation{Joint Space-Science Institute, University of Maryland,
  College Park, MD 20742, USA}

\author[0000-0003-4850-9589]{Martha L.~Boyer}
\affiliation{Space Telescope Science Institute, 3700 San Martin Drive,
  Baltimore, MD 21218, USA}

\author[0000-0002-5189-8004]{Daniela Calzetti}
\affiliation{Department of Astronomy, University of Massachusetts,
  Amherst, MA 01002, USA}

\author[0000-0002-1723-6330]{B.~G.~Elmegreen}
\affiliation{Katonah, NY 10536, USA}

\author[0000-0003-4224-6829]{Brandt A.~L.~Gaches}
\affiliation{Department of Space, Earth and Environment, Chalmers
  University of Technology, Gothenburg SE-412 96, Sweden}

\author[0000-0001-5340-6774]{Karl~D.~Gordon}
\affiliation{Space Telescope Science Institute, 3700 San Martin Drive,
  Baltimore, MD 21218, USA}

\author[0000-0001-9162-2371]{L.~K.~Hunt}
\affiliation{INAF - Osservatorio Astrofisico di Arcetri,
  Largo E. Fermi 5, 50125 Firenze, Italy}

 \author[0000-0001-5448-1821]{R.~C.~Kennicutt}
\affiliation{Steward Observatory, University of Arizona, Tucson, AZ
  85721-0065, USA}
\affiliation{Mitchell Institute for Fundamental Physics and Astronomy,
  Texas A\&M University, College Station, TX 77843-4242, USA}

\author[0000-0002-0560-3172]{Ralf S.~Klessen}
\affiliation{Universit\"{a}t Heidelberg, Zentrum f\"{u}r Astronomie,
  Institut f\"{u}r Theoretische Astrophysik, Albert-Ueberle-Str.\ 2,
  69120 Heidelberg, Germany}
\affiliation{Universit\"{a}t Heidelberg, Interdisziplin\"{a}res
  Zentrum f\"{u}r Wissenschaftliches Rechnen, Im Neuenheimer Feld 225,
  69120 Heidelberg, Germany}
\affiliation{Harvard-Smithsonian Center for Astrophysics, 60 Garden
  Street, Cambridge, MA 02138, U.S.A.}
\affiliation{Elizabeth S. and Richard M. Cashin Fellow at the
  Radcliffe Institute for Advanced Studies at Harvard University, 10
  Garden Street, Cambridge, MA 02138, U.S.A.}

\author[0000-0001-8490-6632]{Thomas S.-Y. Lai}
\affiliation{IPAC, California Institute of Technology, 1200
  E. California Blvd., Pasadena, CA 91125, USA}

\author[0000-0002-2545-1700]{Adam K.~Leroy}
\affiliation{Department of Astronomy, The Ohio State University, 140
  West 18th Avenue, Columbus, OH 43210, USA}
\affiliation{Center for Cosmology and Astroparticle Physics (CCAPP),
  191 West Woodruff Avenue, Columbus, OH 43210, USA}

\author[0000-0002-1000-6081]{Sean T.~Linden}
 \affiliation{Steward Observatory, University of Arizona, 933 N Cherry
  Avenue, Tucson, AZ 85721, USA}

\author[0000-0002-8222-8986]{Alex Pedrini}
\affiliation{Department of Astronomy, Oskar Klein center, Stockholm
  University, AlbaNova, SE-106 91 Stockholm, Sweden}

\author[0000-0002-0361-8223]{Noah S.~J.~Rogers}
\affiliation{Department of Physics and Astronomy, Northwestern
  University, 2145 Sheridan Road, Evanston, IL 60208, USA}
\affiliation{Center for Interdisciplinary Exploration and Research in
  Astrophysics (CIERA), Northwestern University, 1800 Sherman Avenue,
  Evanston, IL 60201, USA}

\author[0000-0001-6326-7069]{Julia C.~Roman-Duval}
\affiliation{Space Telescope Science Institute, 3700 San Martin Drive,
  Baltimore, MD 21218, USA}

\author[0000-0002-3933-7677]{Eva Schinnerer}
\affiliation{Max Planck Institut f\"ur Astronomie, K\"onigstuhl 17,
    D-69117, Heidelberg, Germany}

\author[0000-0003-0605-8732]{Evan B.~Skillman}
\affiliation{University of Minnesota, Minnesota Institute for
  Astrophysics, School of Physics and Astronomy, 116 Church Street
  S.E., Minneapolis, MN 55455, USA}

\author[0000-0003-4793-7880]{Fabian Walter}
\affiliation{Max Planck Institut f\"ur Astronomie, K\"onigstuhl 17,
    D-69117, Heidelberg, Germany}

\author[0009-0005-8923-558X]{Tony D.~Weinbeck}
\affiliation{Department of Physics and Astronomy, University of
  Wyoming, Laramie, WY 82071, USA}

\author[0000-0002-7502-0597]{Benjamin~F.~Williams}
\affiliation{Astronomy Department, University of Washington, Seattle,
  WA 98195, USA}

\email{draine@astro.princeton.edu}

\begin{abstract}

Deuteration of hydrocarbon material, including polycyclic aromatic
hydrocarbons (PAHs), has been proposed to account for the low
gas-phase abundances of D in the interstellar medium (ISM).  JWST
spectra of four star-forming regions in M51 show an emission feature,
with central wavelength $\sim$$4.647\micron$ and FWHM $0.0265\micron$,
corresponding to the C--D stretching mode in aliphatic hydrocarbons.
The emitting aliphatic material is estimated to have ${\rm D/H})_{\rm
  aliph}\approx 0.17\pm0.02$ -- a factor $\sim$$10^4$ enrichment
relative to the overall ISM.  On $\sim$$50\,$pc scales, deuteration
levels toward four \ion{H}{2} regions in M51 are 2-3 times higher than
in the Orion Bar photodissociation region (PDR), with implications for
the processes responsible for the formation and evolution of
hydrocarbon nanoparticles, including PAHs.  The deuteration of the
aliphatic material is found to anticorrelate with helium ionization in
the associated \ion{H}{2}, suggesting that harsh FUV radiation may act
to lower the deuteration of aliphatics in PDRs near massive stars.  No
evidence is found for deuteration of aromatic material, with $({\rm
  D/H})_{\rm arom} \ltsim 0.016$: deuteration of the aliphatic
material exceeds that of the aromatic material by at least a factor
10.  The observed levels of deuteration may account for the depletion
of D observed in the Galactic ISM.  If so, the $4.65\micron$ feature
may be detectable in absorption.

\end{abstract}

\section{Introduction}


The abundance of deuterium provides key insights into many phenomena,
ranging from nucleosynthesis in the early Universe to processes in the
interstellar medium (ISM).  Absorption line studies of gas at high
redshift find $\Da/\Ha \approx (25.5\pm0.3)\ppm$, consistent with the
predictions of Big Bang nucleosynthesis \citep[][and references
  therein]{Yeh+Olive+Fields_2021}.  Absorption line studies of
gas-phase deuterium in the ISM find $(\Da/\Ha)_{\rm gas}$ as high as
$23\ppm$ \citep{Linsky+Draine+Moos+etal_2006}, consistent with the
expected small reduction in $\Da/\Ha$ by the stellar processing
(``astration'') that has enriched the ISM with heavy elements.
However, careful absorption line studies
\citep{Jenkins+Tripp+Wozniak+etal_1999} discovered spatial variability
in $(\Da/\Ha)_\gas$, with very low $(\Da/\Ha)_\gas =
7.4_{-1.3}^{+1.9}\ppm$ in the diffuse gas toward $\delta$Ori.  Low
values of $(\Da/\Ha)_\gas$ have since been observed on many Galactic
sightlines, with $(\Da/\Ha)_\gas\approx 18\ppm$ in the nearby ISM, and
$\sim$$8.6\pm0.8\ppm$ for a number of sightlines with
$5\times10^{20}\ltsim \NH \ltsim 2\times10^{21}\cm^{-2}$
\citep{Linsky+Draine+Moos+etal_2006}.  A recent study by
\citet{Friedman+Chayer+Jenkins+etal_2023} has several sightlines with
$(\Da/\Ha)_\gas < 9\ppm$.  Remarkably, as much as 2/3 of the D is
missing from the gas phase in some diffuse regions.

It has been conjectured that the ``missing'' D -- up to $\sim$$14\ppm$
with respect to H -- has been incorporated into hydrocarbon grains
\citep{Jura_1982,Draine_2004a,Draine_2006b}, presumably including the
polycyclic aromatic hydrocarbon (PAH) population.  Deuterated
hydrocarbons would have emission features in the $4.3-4.7\micron$
range due to the C--D stretching mode
\citep{Allamandola+Tielens+Barker_1989,Allamandola_1993}.

\citet{Verstraete+Puget+Falgarone+etal_1996} found an unidentified
emission band at $4.65\micron$ in the ISO-SWS spectrum of the M17
photodissociation region (PDR), and
\citet{Peeters+Allamandola+Bauschlicher+etal_2004} reported emission
bands at $4.4\micron$ (present at $1.9\sigma$ level) and $4.65\micron$
(present at $4.4\sigma$ level) in the ISO-SWS spectrum of the Orion
Bar, identifying them as C--D stretching modes of deuterated PAHs.

AKARI spectra of a molecular cloud in the LMC showed an unidentified
dust emission feature at $4.65\micron$
\citep{Boulanger+Onaka+Pilleri+Joblin_2011}.
\citet{Doney+Candian+Mori+etal_2016} used AKARI spectra of Galactic
\ion{H}{2} regions to search for deuterated PAHs.  In 6 of 53 targets,
features were seen (at 4.63$\micron$ and $4.75\micron$) that were
attributed to aliphatic C--D stretching modes in PAHs, but
interpretation was difficult due to blending with strong emission
lines observed with the limited spectral resolution $R\approx 100$ of
AKARI.  \citet{Onaka+Sakon+Shimonishi_2022} reported an emission
feature at $4.4\micron$, which they attributed to the aromatic C--D
stretch, in the AKARI spectrum of a massive young stellar object.

\citet{Boersma+Allamandola+Esposito+etal_2023} analyzed JWST spectra
of a number of Galactic sources, confirming the $4.65\micron$ emission
feature in M17, with $\FWHM\approx 0.02\micron$.
\citet{Peeters+Habart+Berne+etal_2024} verified this feature in JWST
spectra of the Orion Bar PDR, with central wavelength
$\lambda_0\approx4.645\micron$,
and reported tentative detection of emission at $4.35\micron$
that they attributed to aromatic C--D.

The unprecedented sensitivity and excellent spectral resolution of
JWST \citep{Gardner+Mather+Abbott+etal_2023} enable us to search in
the Whirlpool galaxy (M51) for evidence that deuterium is sequestered
in hydrocarbon nanoparticles.  Here we report spectra\footnote{%
   All wavelengths in this paper are rest frame, \emph{in vacuo}.}
of four star-forming regions.  All show an emission feature near
$4.65\micron$, attributable to the aliphatic C--D stretch.  After
correcting for the complex gas-phase emission spectrum in this
wavelength range, we find that all four regions are well fitted by a
common band profile, with central wavelength
$\lambda_0\approx4.647\micron$ and $\FWHM=0.0265\micron$.  This
profile is consistent with observations of the Orion Bar.  Relative to
the $3.4\micron$ emission feature, the aliphatic C--D feature is $2-3$
times stronger in M51 than in the Orion Bar, indicating a higher
degree of deuteration.

Our spectra show no evidence of any feature near $4.40\micron$ as
would be expected for the aromatic C--D stretch, nor do we see evidence
of a suggested aliphatic feature near $4.75\micron$.

Our observations are reported in Section \ref{sec:observations}.  The
effects of deuteration are reviewed in Section \ref{sec:deuteration}.
Section \ref{sec:modeling} describes modeling of the gas-phase
emission (and absorption) that must be allowed for to extract the
emission feature.  Our results are presented and discussed in Section
\ref{sec:discussion}, and summarized in Section \ref{sec:summary}.

\section{\label{sec:observations}Observations}
\subsection{M51}

\begin{figure}
\begin{center}
\includegraphics[angle=0,width=18cm,
               clip=true,trim=0.0cm 0.0cm 0.0cm 0.0cm]%
{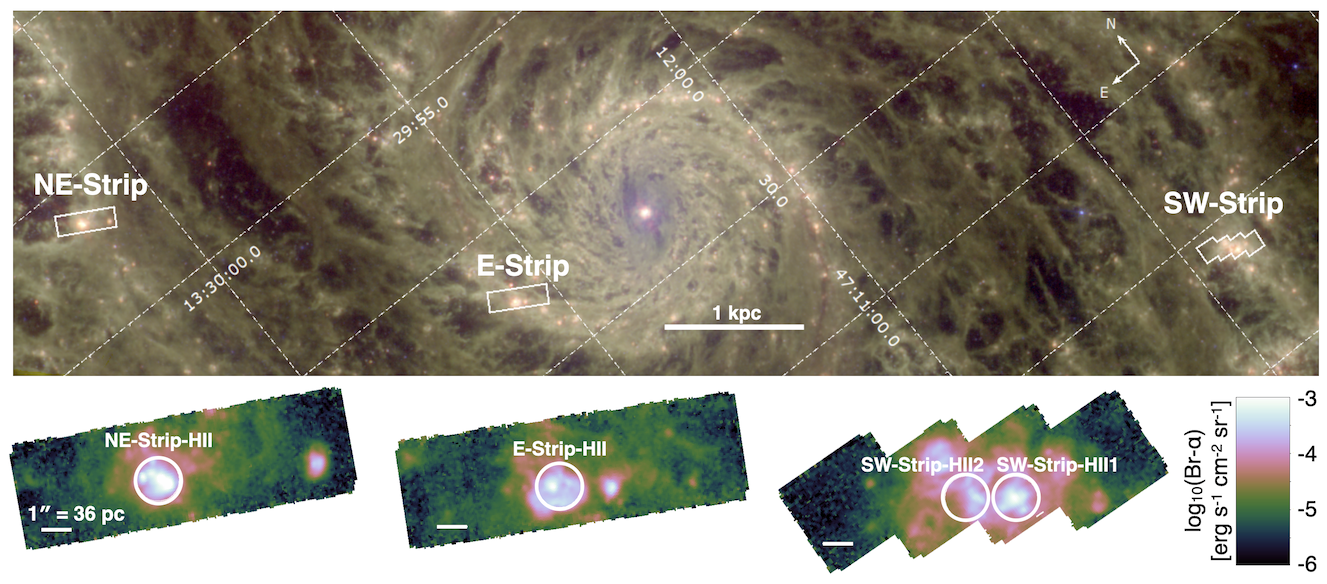} 
\caption{\label{fig:m51coverage}\footnotesize Image of a portion of
  M51; RGB=F1800W/F1130W/F560W ($18\micron$, $11.3\micron$,
  5.6$\micron$).  Areas where we have obtained complete $1-28\micron$
  spectroscopy are indicated.  Expanded cutouts are Br$\alpha$
  ($4.05\micron$) images of these areas.  Spectra have been extracted
  for the four $r=0.75\arcsec$ circular regions shown in the cutouts.
  }
\end{center}
\end{figure}
\begin{figure}
\begin{center}
\includegraphics[angle=0,width=10cm,
               clip=true,trim=0.5cm 5.0cm 0.0cm 4.3cm]
{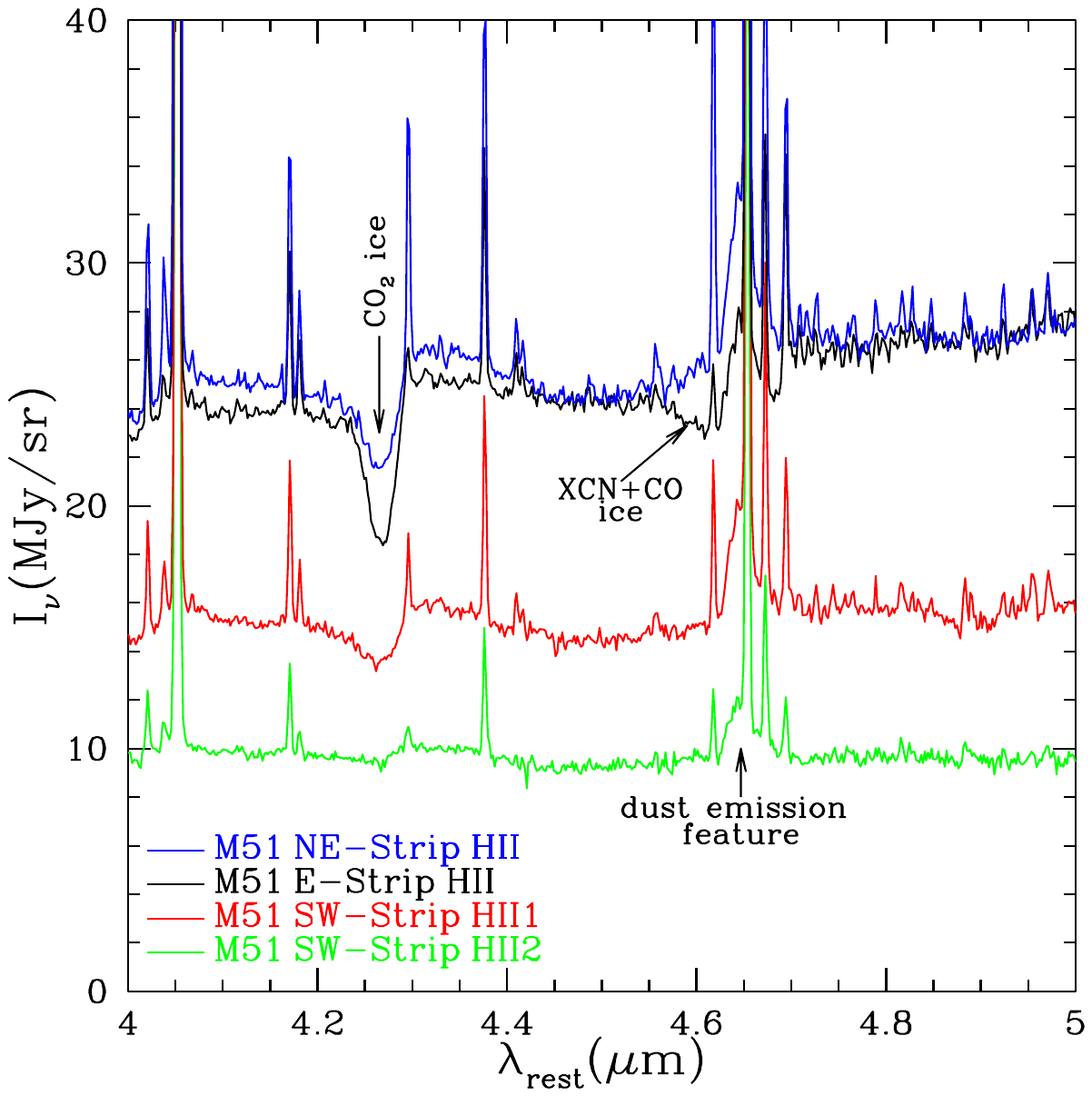}
\caption{\label{fig:m51_4-5}\footnotesize $4-5\micron$ spectra
  ($R\approx 1100$) of four star-forming regions in M\,51 (see text).
  Observed wavelengths have been corrected for redshift (see Table
  \ref{tab:params} for adopted radial velocities).  Three of the
  spectra show CO$_2$ ice absorption (strongest in E-Strip
  \ion{H}{2}).  The E-Strip \ion{H}{2} spectrum also shows
  4.56-4.69$\micron$ XCN+CO ice absorption.  A dust emission feature near
  $4.65\micron$ is present in all spectra.  
  }
\end{center}
\end{figure}

Three regions of M51 have been spectroscopically mapped from $1-28$
\micron\ as part of the Cycle 2 JWST M51 Treasury (GO 3435) using JWST
NIRSpec \citep{Jakobsen+Ferruit+AlvesdeOliveira+etal_2022,
  Boker+Arribas+Lutzgendorf+etal_2022} and MIRI-MRS
\citep{Argyriou+Glasse+Law+etal_2023}.  The NIRSpec observations used
the medium resolution grating/filter combinations G140M/F100LP,
G235M/F170LP, and G395M/F290LP to obtain spectral resolution $R\sim
1000$.  Each region was covered with four individual pointings to
create a small strip that cuts across the spiral arm, with four dither
positions per pointing.  To remove stray light from open Micro-Shutter
Assembly (MSA) shutters, we obtained a ``leakcal'' observation at
every dither position for NIRSpec.

A full description of the observational program and data reduction
will be presented in an upcoming paper (D.A. Dale et al.\ 2025, in
preparation).  In brief, we downloaded the \texttt{uncal} files from the
Mikulski Archive for Space Telescopes (MAST) and reduced them with the
JWST pipeline.  For NIRSpec, we used version 1.16.1.dev14 of the
pipeline (which allows for $1/f$ noise removal from both the science
and leakcal observations) with CRDS \texttt{jwst\_1293.pmap}. We
processed the observations through the Detector1 pipeline, including
$1/f$ noise correction with the \texttt{clean\_flicker\_noise}
step. We then processed the data through the Spec2 and Spec3 pipeline
stages, including the ``leakcal'' subtraction, and created drizzled
cubes with $0\farcs05$ pixels for each grating/filter combination.

We extracted spectra in four $r=0\farcs75$ ($\sim$\,27\,pc at the
7.5\,Mpc distance of M\,51) circular apertures targeting \ion{H}{2}
regions, shown in Figure~\ref{fig:m51coverage}, from each individual
cube, propagating uncertainties from the pipeline-generated error
cube.  The $r=0\farcs75$ size is large compared to the point-spread
function at $\lambda<5\micron$, so we have not convolved the cubes to
a matched angular resolution. To stitch the spectra from each cube
together, we calculated additive offsets (typically
$\sim$$0.1\MJy\sr^{-1}$) between the segments in their overlapping
spectral regions and tied all orders to the G140M spectrum.  In these
bright star-forming regions, the results are not sensitive to the
stitching approach.

The $4-5\micron$ spectra are shown in Figure \ref{fig:m51_4-5}.
Strong emission lines are present, including H and He recombination
lines.  Three of the four spectra show clear evidence of absorption
near $4.26\micron$ due to CO$_2$ ice.  The E-Strip \ion{H}{2} spectrum
also shows an absorption feature near $4.6\micron$ attributed to
``XCN'' ice.  All four spectra show a dust emission feature near
4.65$\micron$, but the spectra are complicated by strong emission
lines: H\,7-5 Pf\,$\beta\,4.654\micron$, multiplets from He\,I
$1s7\ell\rightarrow1s5\ell^{\prime}$ transitions (e.g.,
1s7d $^3{\rm D}_3$$\rightarrow$ 1s5f $^3{\rm  F}_4^{\rm o}$ $4.642\micron$),
and multiplets from He\,I $1s5\ell\!\rightarrow\! 1s4p$ transitions
(e.g.,
1s5s $^3{\rm S}_1$$\rightarrow$ 1s4p $^3{\rm P}_2^{\rm o}$ $4.695\micron$).

\subsection{Orion Bar and M17}

For comparison with M51, NIRSpec spectra (high resolution) of the
Orion Bar \citep[``atomic PDR'' and
  ``DF1'';][]{Peeters+Habart+Berne+etal_2024} were obtained from
\url{https://cdsarc.cds.unistra.fr/viz-bin/cat/J/A+A/685/A74}, and
medium resolution NIRSpec spectra \newtext{(x1d full aperture
  extractions)} of M17 \citep[``M17-PDR'' and
  ``M17B-PDR'';][]{Boersma+Allamandola+Esposito+etal_2023} were
obtained from MAST \url{https://mast.stsci.edu}.

\section{\label{sec:deuteration} Deuteration of Hydrocarbons}

C--D and C--H have identical interatomic potentials.  Because of the
mass difference, the vibrational frequencies for C--D stretching and
bending modes are lower than for C--H.  For free-flying CD and CH
molecules, the difference in reduced masses gives
$\lambda_\CD/\lambda_\CH \approx \sqrt{13/7} = 1.363$.  With the
observed aromatic C--H stretch in PAHs at
$\lambda_\aromCH=3.29\micron$, and the C--H stretch in aliphatic
material at $\lambda_\aliphCH\approx3.42\micron$, the deuterated
counterparts are expected to be near $\lambda_\aromCD\approx
4.48\micron$ and $\lambda_\aliphCD \approx 4.66\micron$.  More
detailed studies estimate $\lambda_\aromCD\approx4.40\micron$ and
$\lambda_\aliphCD\approx4.65\micron$ \citep[][V.~J.~Esposito 2024,
  private communication]{Bauschlicher+Langhoff+Sandford+Hudgins_1997,
  Hudgins+Bauschlicher+Sandford_2004, Yang+Li+Glaser_2020b,
  Yang+Li_2023a}.

Because the zero-point energy in stretching and bending modes is lower
for CD than for CH, the bond energy is greater by $\sim$$0.083\eV$
\citep{Draine_2006b}.  With CD energetically favored over CH,
one might expect interstellar hydrocarbons to have D/H ratios larger
than $(\Da/\Ha)_\gas$.

\section{\label{sec:modeling} Modeling the Observed Emission}

\subsection{Recombination Lines and Reddening}

Our goal is to examine the spectra for evidence of emission from
deuterated hydrocarbons in M51.  In addition to our spectra of four
star-forming regions in M51, for comparison we apply the same analysis
to spectra of the ``atomic PDR'' and ``Dissociation Front 1''
(``DF1'') positions in the Orion Bar PDR
\citep{Peeters+Habart+Berne+etal_2024}, and two positions in the M17
PDR \citep{Boersma+Allamandola+Esposito+etal_2023}.

We model the observed spectra as line emission plus dust emission
attenuated by a foreground screen of dust with ices, and gas-phase CO:
\beq \label{eq:I_nu}
I_\nu(\lambda) = I_\nu^{\rm line}
\exp\left[-(\tau_\lambda^\dust+\tau_\lambda^{\rm ice})\right]
+
\left(I_\nu^\cont+I_\nu^{\rm CD}\right)
\exp\left[-(\tau_\lambda^\dust+
            \tau_\lambda^{\rm ice}+\tau_\lambda^{\rm CO})\right]
~~~,
\eeq
where the line emission $I_\nu^{\rm line}$ and gas-phase CO absorption
$\tau_\lambda^{\rm CO}$ are obtained from a model, $I_\nu^{\rm CD}$ is
the modeled aliphatic C--D emission feature discussed below, and
$I_\nu^\cont\exp[-\tau_\lambda^\dust]$ is the \emph{observed}
continuum (including PAH features, \emph{except} for the aliphatic
C--D feature $I_\nu^{\rm CD}$) interpolated over wavelengths where ice
absorption is present.  $I_\nu^{\rm line}$ includes emission from
$v=1$ CO (Section \ref{sec:other emission lines}).  We assume that the
hot $v=1$ CO is not at the same velocity as the bulk of the $v=0$ CO,
so that they are not significantly absorbed by $v=0$ CO.
 
The H and He recombination lines are modeled using case B
recombination calculations for \ion{H}{2} \citep{Storey+Hummer_1995}
and \ion{He}{2} \citep{DelZanna+Storey_2022}. The
\ion{He}{1}\,10833$\Angstrom$ triplet is assumed to be enhanced by a
factor $\sim$$1.2$ due to radiative transfer effects (B.T.~Draine et al.\ 2025, in preparation).  For the four M51 \ion{H}{2} regions we take the
electron temperature $T_e=6000\K$ \citep[as determined for these
  regions by][]{Croxall+Pogge+Berg+etal_2015}, and electron density
$n_e=10^3\cm^{-3}$.  For the ionized gas in the Orion Bar we take
$T_e=9000\K$ and $n_e=3000\cm^{-3}$
\citep{Pogge+Owen+Atwood_1992,Blagrave+Martin+Rubin+etal_2007}.  For
the ionized gas in M17 we take $n_e=500\cm^{-3}$ and $T_e\approx
8500\K$ \citep{Garcia-Rojas+Esteban+Peimbert+etal_2007}.\footnote{%
For most of the H and He recombination lines, line ratios are
insensitive to reasonable variations in $n_e$ and $T_e$.}
For \ion{H}{1} we include recombination lines from levels up to
$n=49$.  Simple Gaussian line profiles are assumed for all lines,
using the published $R\equiv\lambda/\FWHM$ for NIRSpec
\citep{Jakobsen+Ferruit+AlvesdeOliveira+etal_2022}.\footnote{%
%
Our M51 spectra and the M17 spectra were obtained with the NIRSpec
medium resolution mode.  The Orion Bar spectra
\citep{Peeters+Habart+Berne+etal_2024} were obtained with the high
resolution mode.}

\begin{figure}
\begin{center}
\includegraphics[angle=0,width=\fwidth,
                 clip=true,trim=0.5cm 5.0cm 0.0cm 3.5cm]
{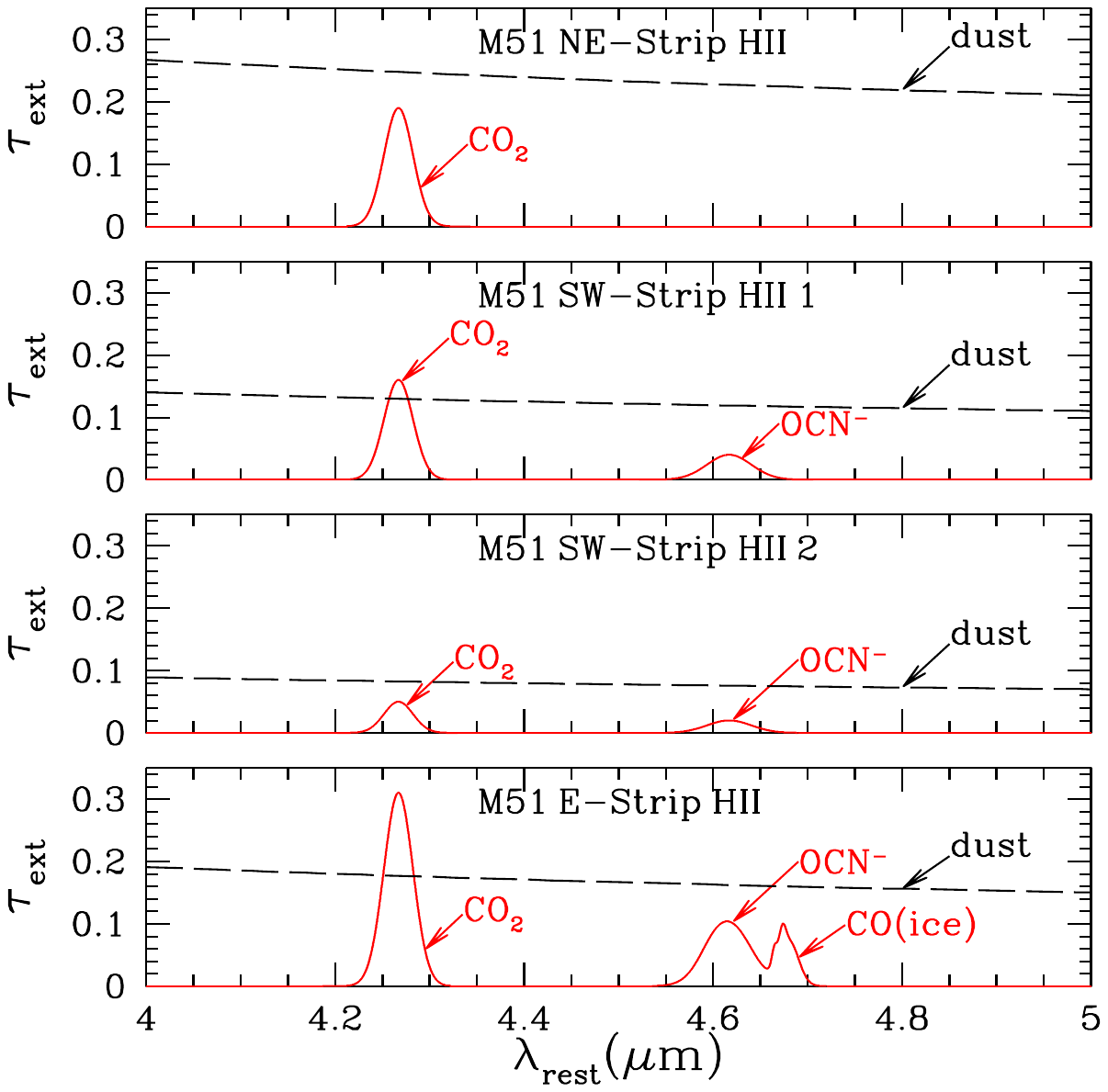}
\includegraphics[angle=0,width=\fwidth,
                 clip=true,trim=0.5cm 5.0cm 0.0cm 3.5cm]
{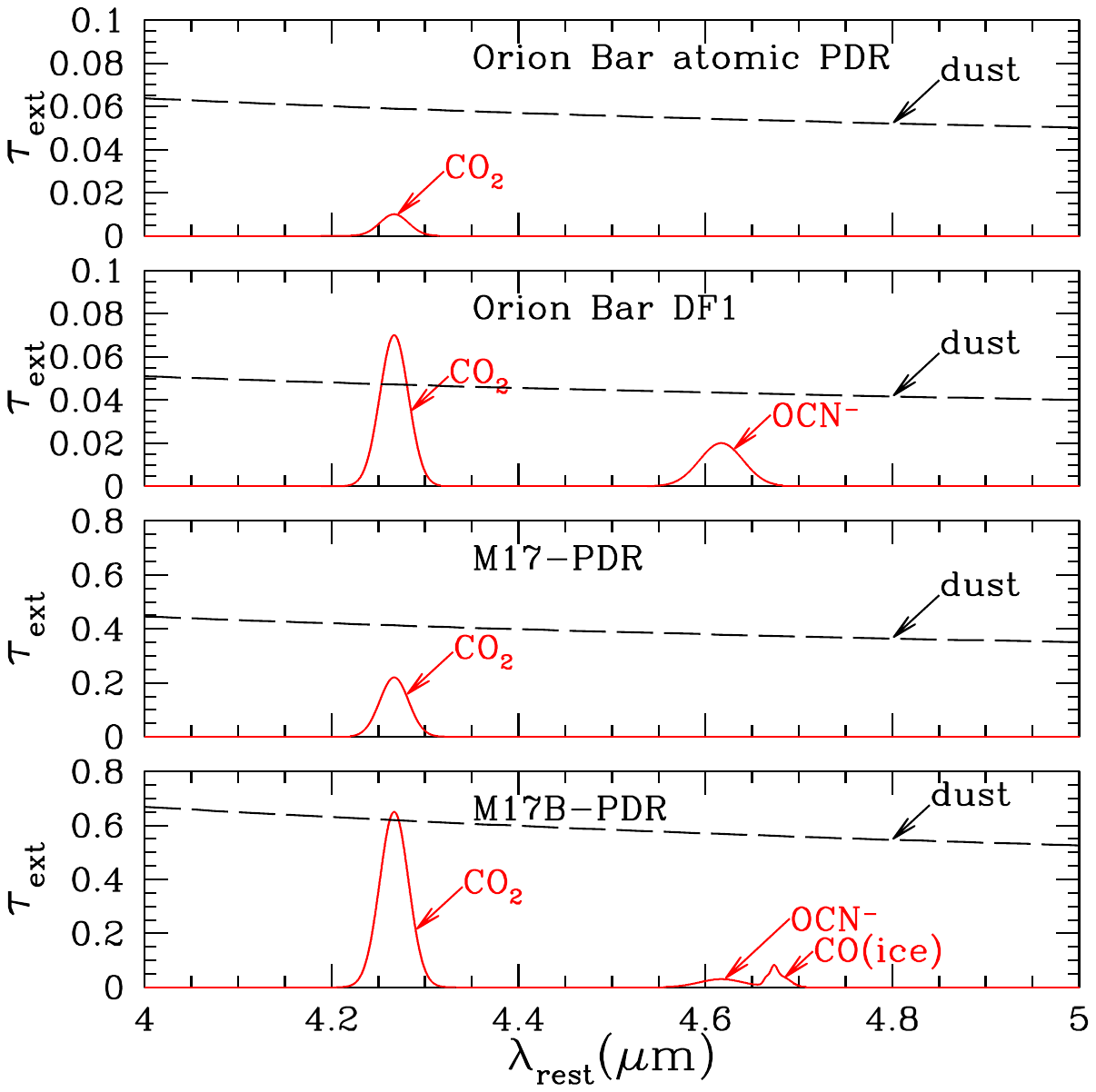}
\caption{\label{fig:tau}\footnotesize Estimated extinction due to dust
  (dashed curves) and ices (red curves) over the $4-5\mu$m range for
  the four regions in M51 (see text).  
  }
\end{center}
\end{figure}

\begin{figure}
\begin{center}
\includegraphics[angle=270,width=18.0cm,
                 clip=true,trim=0.5cm 0.5cm 0.0cm 0.5cm]
{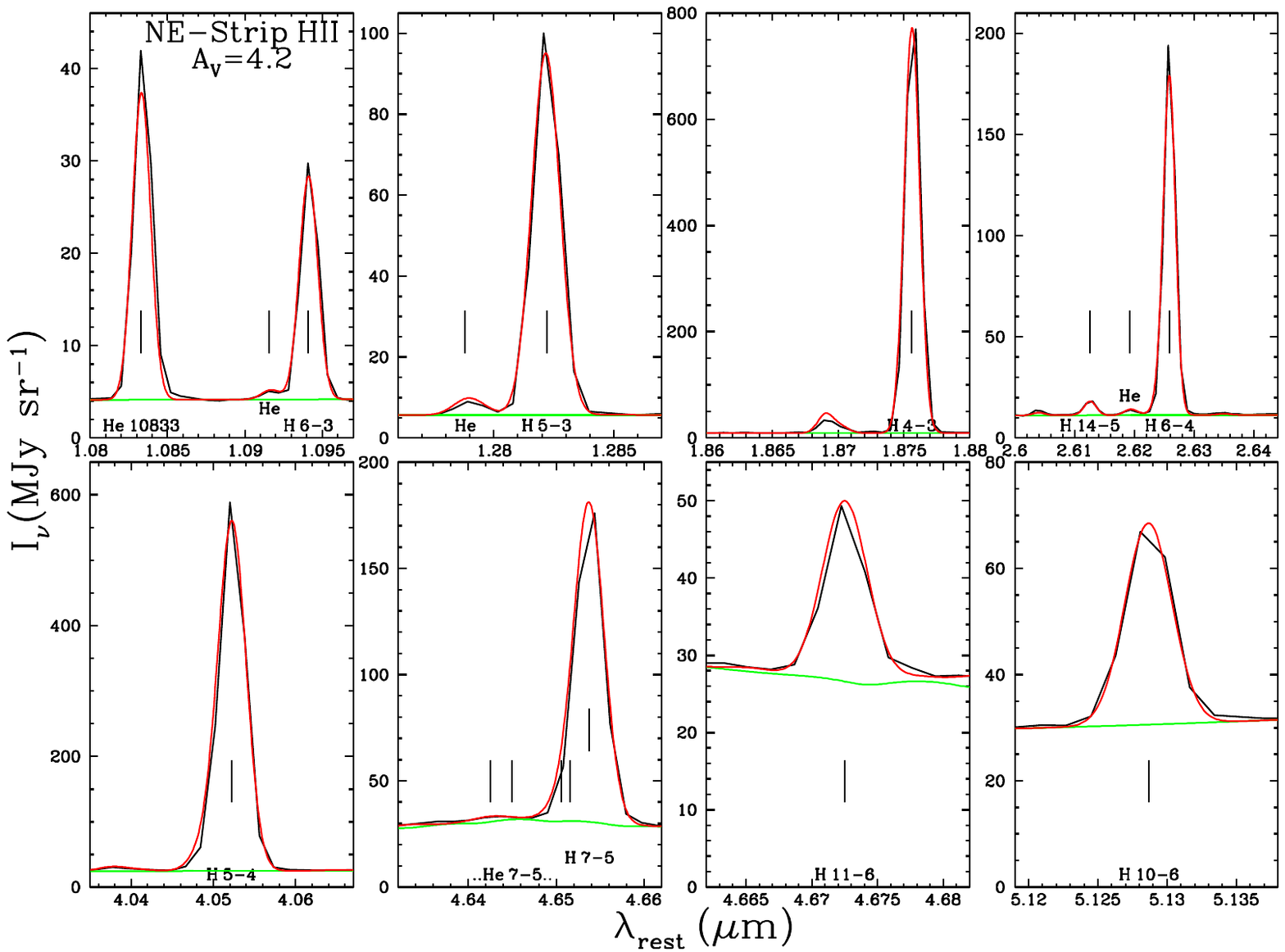}
\caption{\label{fig:H lines}\footnotesize Selected recombination lines
  for NE\,Strip\,\ion{H}{2}.  Black curves: observed spectrum.  Red
  curves: model = continuum (green) + line emission.  The differential
  extinction between $1.094\mu$m (Pa$\gamma$) and $5.129\mu$m
  (Hu$\delta$) is consistent with the adopted
  \citet{Gordon+Clayton+Decleir+etal_2023} $R_V=5.5$ extinction curve
  and $A_V=4.2\,$mag, with differential reddening
  $A_{1.09}-A_{5.13}=0.408A_V = 1.71\,$mag determined to
  $\pm0.10\,$mag.
  }
\end{center}
\end{figure}

We fit emission measures $\EM \equiv \int n_e
n(\Ha^+)dL$ and He ionization $\He^+/\Ha^+$ to reproduce the observed
$1-5\micron$ H and He recombination lines (including \ion{He}{1}
$1.0833\micron$, H\,Pa$\gamma\,1.0941\micron$,
H\,Pa$\alpha\,1.876\micron$, and H\,Br$\alpha\,4.052\micron$) in each
region.  Reddening by dust (needed to reproduce the observed line
ratios) is included, modeled as a uniform foreground screen, using the
$R_V=5.5$ extinction curve from
\citet{Gordon+Clayton+Decleir+etal_2023}.  Table \ref{tab:params}
gives $\EM$, $\He^+/\Ha^+$, and $A_V$ for each region.  Figure
\ref{fig:tau} shows the adopted $\tau_\dust(\lambda)$ for the
four M51 regions in M51.

Figure \ref{fig:H lines} shows selected recombination lines for our
most heavily-reddened M51 source, NE-Strip \ion{H}{2}
($A_V=4.2\,$mag).  Agreement is excellent.  The \ion{He}{1}~7-5
emission near $4.65\micron$ consists of a number of blended lines; the
wavelengths of the strongest lines are shown.  Similar results are
found for the other three regions.  Evidently, the adopted extinction
law works well for $1-5\micron$.

\subsection{\label{sec:ice} Ice Absorption}

Three of the four spectra in Figure \ref{fig:m51_4-5} show strong
absorption near $4.26\micron$ due to CO$_2$ ice
\citep{Gibb+Whittet+Boogert+Tielens_2004}, and the E-Strip \ion{H}{2}
spectrum has additional ice absorption features between
$\sim$$4.55\micron$ and $\sim$$4.70\micron$ due to ``XCN'' \citep[$=$
  OCN$^-$:][]{vanBroekhuizen+Keane+Schutte_2004,
  Oberg+Boogert+Pontoppidan+etal_2011} ice with a possible
contribution from CO ice
\citep{Gibb+Whittet+Boogert+Tielens_2004,Onaka+Sakon+Shimonishi_2022}.
We determine $\tau_{\rm ice}(\lambda)$ by comparing the observed
continuum with a smooth continuum interpolated between $4.1$ and
$4.3\micron$, and between 4.50 and $4.75\micron$ (making allowance for
the dust emission feature $I_\nu^{\rm CD}$ centered at $4.65\micron$).
Our estimate for the optical depth of the CO$_2$, OCN$^-$, and CO ice
features is shown in Figure \ref{fig:tau}.
 
\subsection{\label{sec:other emission lines} Other Emission Lines}

The $4-5\micron$ spectra also include emission lines from $\HH$ and
\ion{K}{3}, which have been added to the model spectrum to reproduce the
observed line strengths (see Table \ref{tab:params}).  For the Orion
Bar positions we also add the \ion{O}{1} multiplet at $4.561\micron$
reported by \citet{Peeters+Habart+Berne+etal_2024}, presumed to be
fluorescent emission.  Our M51 spectra show no trace of this line.

Some of our M51 spectra show CO $v=1\rightarrow0$ emission lines, as
previously seen in the Orion Bar
\citep{Peeters+Habart+Berne+etal_2024}.  Thus, our models include
emission from $v=1$ CO, with rotational states $(v=1,J)$ populated
according to the rotational temperature $T_{{\rm rot,CO},v=1}$ given in
Table \ref{tab:params}.  We use Einstein $A$ coefficients from
\citet{Chandra+Maheshwari+Sharma_1996}.  The CO emission is treated as
optically thin; the lines are assumed to be sufficiently displaced in
velocity to be unaffected by absorption by $v=0$ CO (see below).

\subsection{CO Absorption Lines}

Some of our model spectra (Equation \ref{eq:I_nu}) include
$v=0\!\rightarrow\!1$ absorption by CO, with the $v=0$ rotational
levels assumed to be thermally populated with rotational excitation
temperatures $T_{{\rm rot,CO},v=0}$ (see Table \ref{tab:params}). The
adopted CO column densities $N(\CO)\approx 10^{17}\cm^{-2}$ are
consistent with $n(\CO)/n(\HH)\approx 2\times10^{-4}$ in the molecular
gas along the line of sight, as observed for cold molecular gas in the
Galaxy \citep{Lacy+Sneden+Kim+Jaffe_2017}, although $N(\CO)$ is not
well constrained by the current data.  See the Appendix for further
details.

\subsection{Continuum and Dust Features}

A smoothed continuum (emission from stars, dust, free-bound and
free-free) was estimated for each of the regions, using the observed
spectra between emission lines.  The dust emission includes the
well-known PAH emission features at $3.29$ and $3.4\micron$,
as well as weaker dust emission features and ice absorption features.
The modeled emission and absorption lines are added to this continuum.
As discussed below (Section \ref{sec:aliphCD}), we find that we must
add an emission feature near $4.65\micron$ to reproduce the observed
spectra.

\subsection{Uncertainties}

In our model, all of the emission is assumed to be attenuated by a
foreground screen with a single $A_V$, which is clearly unrealistic.
Because the modeling also involves subjective decisions regarding (1)
placement of continuum levels, (2) estimation of the strength and
wavelength dependence of absorption by ices, and (3) choices regarding
the central wavelength and FWHM of the $4.65\micron$ feature, we do
not attempt rigorous estimation of uncertainties.  The uncertainties
listed for different quantities in Table \ref{tab:params} are our
subjective estimates; they are larger than would be estimated from
statistical uncertainties alone.

\begin{table}
\addtolength{\tabcolsep}{-0.25em}
\newcommand \fnA  {a}
\newcommand \fnB  {b}
\newcommand \fnC  {c}
\newcommand \fnD  {d}
\newcommand \fnE  {e}
\newcommand \fnF  {f}
\newcommand \fnG  {g}
\newcommand \fnH  {h}
\newcommand \fnI  {i}
\newcommand \fnJ  {j}
\begin{center}
\caption{\label{tab:params} Emission Parameters for Modeled Regions
}
{\scriptsize
\hspace*{-2.8cm}\begin{tabular}{c c c c c c c c c}
\hline
       & M51
       & M51
       & M51
       & M51
       & Orion
       & Orion 
       & M17
       & M17-B
\\
       & NE-Strip$^\fnA$
       & SW-Strip,1$^\fnB$
       & SW-Strip,2$^\fnB$
       & E-Strip$^\fnC$
       & Atomic PDR
       & DF1
       & PDR
       & PDR
\\
\hline
$v_r\,(\kms)$$^\fnD$
                 & 420 & 500 & 485 & 500 & 18 & 18 & 25 & 15
\\
$R_{\rm gc} (\kpc)$$^\fnE$ 
                 & 4.13 
                 & 4.25
                 & 4.34
                 & 1.12
                 & 8.2
                 & 8.2
                 & 6.5
                 & 6.5
\\
$n_e (\cm^{-3})$       
                 & $10^3$
                 & $10^3$
                 & $10^3$
                 & $10^3$
                 & $3\times10^3$
                 & $3\times10^3$
                 & $500$
                 & $500$
\\
$T_e (\K)$       
                 & $6\times10^3$
                 & $6\times10^3$
                 & $6\times10^3$
                 & $6\times10^3$
                 & $9\times10^3$
                 & $9\times10^3$
                 & $8\times10^3$
                 & $8\times10^3$
\\
$\EM\,(10^{23}\cm^{-5})$ 
                 & $3.65\pm0.1$
                 & $1.90\pm0.1$
                 & $1.00\pm0.1$
                 & $2.05\pm0.10$
                 & $60.\pm5$
                 & $30.\pm3$
                 & $14.5\pm 1.0$
                 & $120.\pm10$
\\
$\dot{N}_L (10^{48}\s^{-1})$
                 & 3100
                 & 1600
                 &  850
                 & 1700
                 & --
                 & --
                 & --
                 & --
\\
$\He^+/\Ha^+$    & $0.030\!\pm\!0.002$
                 & $0.020\!\pm\!0.002$
                 & $0.023\!\pm\!0.002$
                 & $0.015\!\pm\!0.002$
                 & $0.036\!\pm\!0.002$
                 & $0.036\!\pm\!0.002$
                 & $0.052\!\pm\!0.005$
                 & $0.038\!\pm\!0.004$
\\
$A_V$ (mag)      & $4.2\pm0.2$
                 & $2.2\pm0.2$
                 & $1.4\pm0.2$
                 & $3.0\pm0.2$
                 & $1.0\pm0.2$
                 & $0.8\pm0.2$
                 & $7.0\pm0.5$
                 & $10.5\pm0.5$
\\
$N_{{\rm CO},v=0}$ ($10^{16}\cm^{-2}$)
                 & $10$
                 & $1$
                 & $0.5$
                 & $30$
                 & $0.3$
                 & $0.5$
                 & $10$
                 & $10$
\\
$T_{{\rm rot,CO},v=0}$ (K)
                 & $15$
                 & $15$
                 & $15$
                 & $40$
                 & $15$
                 & $15$
                 & $15$
                 & $15$
\\
$N_{{\rm CO},v=1}$ ($10^6\cm^{-2}$)
                 & $10$
                 & $5$
                 & $2$
                 & $10$
                 & $30$
                 & $30$
                 & $100$
                 & $50$
\\
$T_{{\rm rot,CO},v=1}$ (K)
                 & $50$
                 & $100$
                 & $100$
                 & $100$
                 & $120$
                 & $100$
                 & $100$
                 & $100$
\\
$F$($\HH$$\,4.4096\micron$)$^{\fnF}$ 
                 & $2.0\pm0.5$
                 & $1.1\pm0.5$
                 & $0.25\pm0.25$
                 & $1.2\pm0.6$
                 & $15\pm2$
                 & $32\pm3$
                 & $20\pm2$
                 & $24\pm5$
\\
$F$($\HH$$\,4.4171\micron$)$^{\fnF}$ 
                 & $1.5\pm0.5$
                 & $0.7\pm0.5$
                 & $0.2\pm0.2$
                 & $0.7\pm0.6$
                 & $12\pm2$
                 & $20\pm3$
                 & $16\pm2$
                 & $20\pm4$
\\
$F$(\ion{O}{1}\,4.561$\micron$)$^{\fnF,\fnH}$
                 & $0.7\pm0.7$
                 & $<0.3$
                 & $<0.2$
                 & $<0.3$
                 & $50\pm 5$
                 & $4\pm 1$
                 & $4\pm4$
                 & $12\pm4$
\\
$F$($\HH$$\,4.5757\micron$)$^{\fnF}$  
                 & $0.4\pm0.4$
                 & $0.2\pm0.2$
                 & $<1$
                 & $0.4\pm0.3$
                 & $3\pm2$
                 & $10\pm2$
                 & $4\pm4$
                 & $9\pm2$
\\
$F$([\ion{K}{3}]$4.6180\micron$)$^{\fnF}$
                 & $11.5\pm1.0$
                 & $4.5\pm0.5$
                 & $1.6\pm0.2$
                 & $1.1\pm0.1$
                 & $95\pm10$
                 & $53\pm5$
                 & $45\pm5$
                 & $430\pm40$
\\
$F$(H$_2$$\,4.6947\micron$)$^{\fnF}$ 
                 & $7.0\pm0.7$
                 & $4.3\pm0.5$
                 & $1.5\pm0.3$
                 & $6.0\pm0.5$
                 & $65\pm6$
                 & $155\pm10$
                 & $90\pm5$
                 & $110\pm10$
\\
$B_\aliphCD$ ($\MJy \sr^{-1}$)$^{\fnG}$
                 & $8.5\pm 0.8$         
                 & $6.5\pm 1.0$         
                 & $2.7\pm 0.7$         
                 & $5.0\pm 1.3$         
                 & $75\pm 11$           
                 & $55\pm 8$            
                 & $45\pm9$             
                 & $90\pm18$            
\\
$B_{4.35\mu{\rm m}}$ ($\MJy \sr^{-1}$)$^{\fnG}$
                 & $<1.0$
                 & $<0.7$
                 & $<0.5$
                 & $<1.0$
                 & $<20$
                 & $<10$
                 & $<10$
                 & $<10$
\\
$B_{4.40\mu{\rm m}}$ ($\MJy \sr^{-1}$)$^{\fnG}$
                 & $<1.0$
                 & $<0.5$
                 & $<0.5$
                 & $<1.0$
                 & $<10$
                 & $<10$
                 & $<8$
                 & $<10$
\\
$B_{\rm 4.75\mu{\rm m}}$ ($\MJy \sr^{-1}$)$^{\fnG}$
                 & $<1.0$
                 & $<1.0$
                 & $<0.5$
                 & $<1.5$
                 & $<15$
                 & $<15$
                 & $<8$
                 & $<20$
\\
$F_\aliphCD(4.65\micron)$
                 & $38.\pm 4$                     
                 & $32.\pm 5$                     
                 & $14.\pm 4$                     
                 & $23.\pm 6$                     
                 & $410.\pm 60$ 
                 & $300.\pm 45$ 
                 & $175.\pm 35$                   
                 & $280.\pm 60$ 
\\
$F_\aromCD(4.40\micron)$
                 & $<18$               
                 & $<8$                
                 & $<8$                
                 & $<17$               
                 & $<150$              
                 & $<150$              
                 & $<200$              
                 & $<200$              
\\
$F_\aromCD(4.75\micron)$
                 & $<8$                  
                 & $<9$                  
                 & $<5$                  
                 & $<13$                 
                 & $<140$  
                 & $<140$  
                 & $<55$                 
                 & $<110$                
\\
\hline
$F_\aromCH^{\rm\,clip,corr}(3.29\mu{\rm m})$$^{\fnF,\fnG}$
                 & $1610\pm80$
                 & $915\pm50$
                 & $376\pm20$
                 & $925\pm50$
                 & $36500\pm1100$
                 & $23300\pm700$
                 & $8780\pm440$
                 & $13300\pm670$
\\
$F_\nonaromCH^{\rm\,clip,corr}(3.4\mu{\rm m})$$^{\fnF,\fnG}$
                 & $395\pm40$
                 & $225\pm23$
                 & $87\pm9$
                 & $212\pm21$
                 & $7380\pm370$
                 & $5140\pm260$
                 & $2500\pm250$
                 & $3860\pm390$
\\
$F_\aliphCD^{\rm\,corr}(4.65\mu{\rm m})$$^{\fnF,\fnG}$
                 & $49\pm5$            
                 & $38\pm6$            
                 & $16\pm4$            
                 & $43\pm7$            
                 & $434\pm65$          
                 & $318\pm48$          
                 & $260\pm50$
                 & $520\pm100$
\\
$F_\aromCD^{\rm\,corr}(4.40\micron)$$^{\fnF,\fnG}$
                 & $<14$            
                 & $<7$             
                 & $<7$             
                 & $<14$            
                 & $<140$           
                 & $<140$           
                 & $<110$           
                 & $<140$           
\\
$F_{\rm nitrile CN}^{\rm\,corr}(4.35\micron)$$^{\fnF,\fnG,\fnI}$
                 & $<15$            
                 & $<10$            
                 & $<7$             
                 & $<15$            
                 & $<300$           
                 & $<150$           
                 & $<150$           
                 & $<150$           
\\
$F_\aromCD^{\rm\,corr}(4.75\micron)$$^{\fnF,\fnG,\fnJ}$
                 & $<10$            
                 & $<10$            
                 & $<5$             
                 & $<15$            
                 & $<150$           
                 & $<150$           
                 & $<80$            
                 & $<200$           
\\
$F_\nonaromCH^{\rm\,clip,corr}/F_\aromCH^{\rm clip,corr}$$^\fnF$
                 & 0.246  
                 & 0.246
                 & 0.232
                 & 0.229
                 & 0.202  
                 & 0.221
                 & 0.284
                 & 0.290
\\
$F_\aliphCD^{\rm\,corr}/F_\nonaromCH^{\rm\,clip,corr}$$^{\fnF}$
                 & 0.12   
                 & 0.17   
                 & 0.18   
                 & 0.14   
                 & 0.059  
                 & 0.062  
                 & 0.10   
                 & 0.13   
\\
$F_\aromCD^{\rm\,corr}/F_\aromCH^{\rm\,clip,corr}$$^{\fnF}$
                 & $<0.009$   
                 & $<0.008$   
                 & $<0.019$   
                 & $<0.015$   
                 & $<0.004$   
                 & $<0.006$   
                 & $<0.013$   
                 & $<0.010$   
\\
$F_{\rm nitrile CN}^{\rm\,corr}/F_\aromCH^{\rm\,clip,corr}$$^{\fnF}$
                 & $<0.009$   
                 & $<0.011$   
                 & $<0.019$   
                 & $<0.015$   
                 & $<0.008$   
                 & $<0.006$   
                 & $<0.016$   
                 & $<0.011$   
\\
$F_{4.75\mu{\rm m}}^{\rm\,corr}/F_\aromCH^{\rm\,clip,corr}$$^{\fnF}$
                 & $<0.006$ 
                 & $<0.011$ 
                 & $<0.013$ 
                 & $<0.016$ 
                 & $<0.004$ 
                 & $<0.006$ 
                 & $<0.009$ 
                 & $<0.015$ 
\\
\hline
\multicolumn{7}{l}{$\fnA$ NE-Strip\,\ion{H}{2} = NGC5194+91.0+69.0
  \citep{Croxall+Pogge+Berg+etal_2015}}
\\
\multicolumn{7}{l}{$\fnB$ SW-Strip\,\ion{H}{2}-1 and 
  \ion{H}{2}-2 = NGC 5194-86.5-79.4 \citep{Croxall+Pogge+Berg+etal_2015}}
\\
\multicolumn{7}{l}{$\fnC$
  E-Strip\,\ion{H}{2} = NGC5194+30.2+2.2
  \citep{Croxall+Pogge+Berg+etal_2015}}
\\
 \multicolumn{7}{l}{$\fnD$ Heliocentric radial velocity.}
\\
\multicolumn{7}{l}{$\fnE$ Distance from galaxy center.}
\\
\multicolumn{7}{l}{$\fnF$ Surface brightness of line or feature
  ($10^{-6}\erg\cm^{-2}\s^{-1}\sr^{-1}$)} \\
\multicolumn{7}{l}{$\fnG$ Extinction-corrected.}\\
\multicolumn{7}{l}{$\fnH$ \ion{O}{1}~$2p^34p\,\,^3{\rm P}_J\!
                    \rightarrow\!2p^33d\,\,^3{\rm D}_{J^\prime}^{\rm o}$.}
\\
\multicolumn{7}{l}{$\fnI$ Assuming $\FWHM\ltsim 0.06\micron$.}
\\
\multicolumn{7}{l}{$\fnJ$ Assuming $\FWHM=0.047\micron$
                   \citep{Doney+Candian+Mori+etal_2016}.}
\end{tabular}
}
\end{center}
\end{table}

\section{\label{sec:discussion} Results and Discussion}

\subsection{The Selected Emission Regions}

The areas in M51 selected here all include giant \ion{H}{2} regions.
From the hydrogen recombination lines, we infer the rate of hydrogen
photoionizations,
\beq
\dot{N}_L =
\alpha_B\, \EM\, D^2 \,\Omega = 
8.5\times 10^{50}
\left(\frac{\alpha_B}{3.8\times10^{-13}\cm^3\s^{-1}}\right) 
\left(\frac{\EM}{10^{23}\cm^{-5}}\right)
\left(\frac{D}{7.5\Mpc}\right)^2 \s^{-1}
\eeq
for each region, where $\EM$ is the emission measure averaged over the
extraction area with solid angle $\Omega=4.15\times10^{-11}\sr$,
$\alpha_B$ is the ``case B'' recombination rate coefficient
($\alpha_B\approx 3.8\times10^{-13}\cm^3\s^{-1}$ for
$T_e\approx6000\K$), and $D=7.5\Mpc$
\citep{Csornyei+Anderson+Vogl+etal_2023} is the distance of M51. The
inferred $\dot{N}_L$ range from $9\times10^{50}\s^{-1}$ for
E-Strip\,\ion{H}{2} to $3.1\times10^{51}\s^{-1}$ for NE-Strip
\ion{H}{2}.

The $\He^+/\Ha^+$ values of $0.015-0.030$ inferred from the
\ion{He}{1} recombination lines (see Table \ref{tab:params}) from the
four regions in M51 correspond to ionization by stars of spectral type
between O8.5V and O9V, with Lyman continuum photon emission rate
$Q_0\approx 10^{48.1}\s^{-1}$ \citep{Martins+Schaerer+Hillier_2005}.
Thus, each of the four regions contains $\sim$$10^3$ O stars (and many
more B stars).

Molecular cloud material is abundant in each of these star-forming
regions, given the CO$_2$ ice absorption features in all four spectra
(see Figure \ref{fig:m51_4-5}), strong $\HH$ rovibrational emission
lines, and evidence of gas-phase CO absorption and emission lines in
some of the spectra (see below).  Much of the stellar radiation is
presumably reprocessed in photodissociation regions (PDRs) at the
boundaries between \ion{H}{2} regions and molecular clouds.

\subsection{\label{sec:CH Stretch} The C--H Stretch Emission Features}

\begin{figure}
\begin{center}
\includegraphics[angle=0,width=\fwidth,
                 clip=true,trim=0.5cm 5.0cm 0.0cm 2.0cm]
{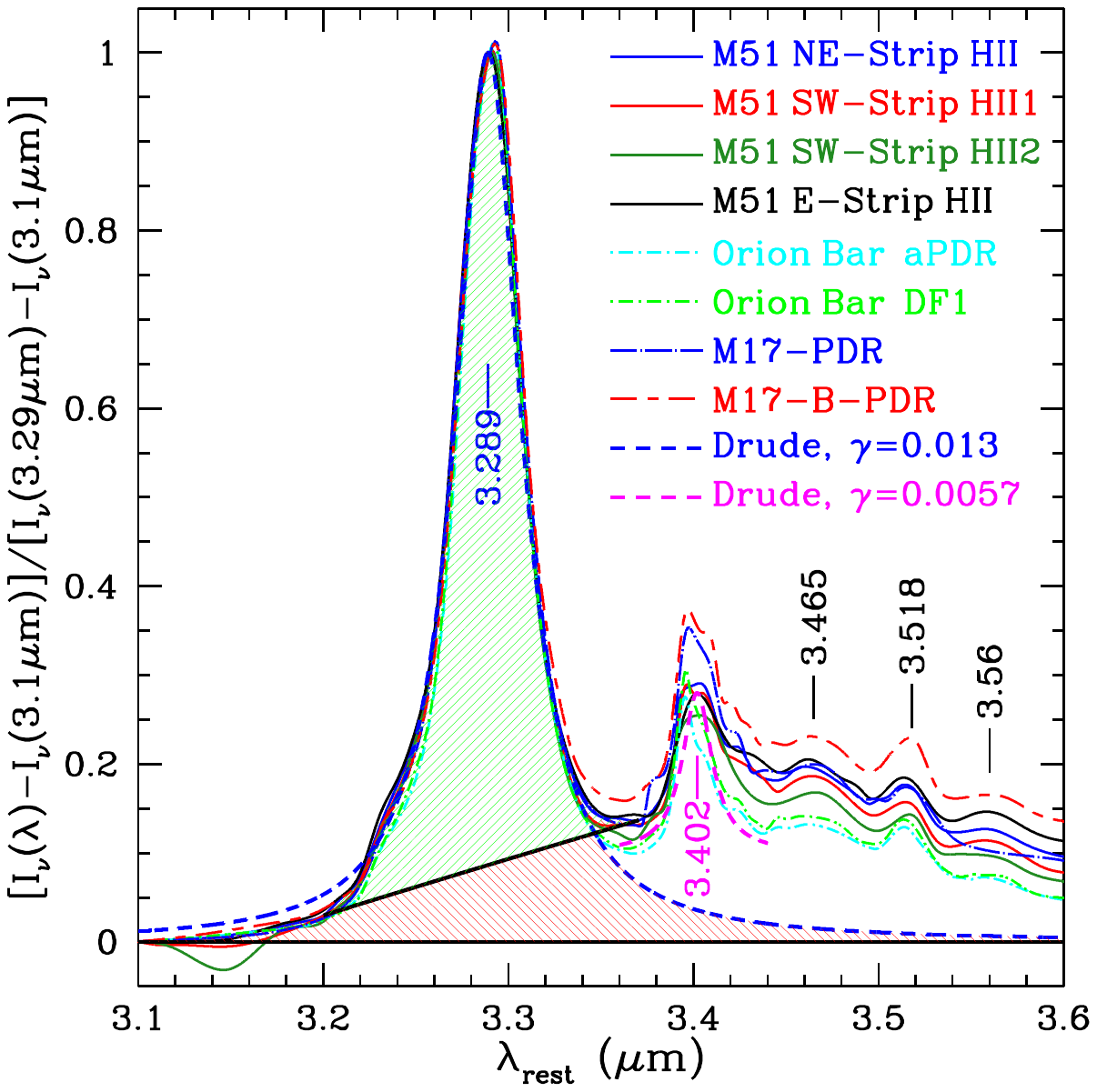}
\caption{\label{fig:33+34}\footnotesize Comparison of the
  $3.1-3.65\micron$ emission in the different regions.  The
  $3.29\micron$ feature profile is nearly invariant.  The
  $3.37$ to $3.60\micron$ emission shows regional variations in strength
  (relative to the $3.29\micron$ feature), and in spectral details
  (see text). Dashed curves: Drude profiles for the $3.29$ and
  $3.40\micron$ components (see text).  The green shaded area shows
  the power included in $F_\aromCH^{\rm\,clip}$; the red shaded area
  shows the power in the Drude profile that is missed in
  $F_\aromCH^{\rm\,clip}$.  We find
  $F_\aromCH^{\rm\,clip}/F_\aromCH\approx 0.71$.}
\end{center}
\end{figure}
\begin{figure}[b]
\begin{center}
\includegraphics[angle=0,width=\fwidth,
                 clip=true,trim=0.5cm 5.0cm 0.0cm 2.0cm]
{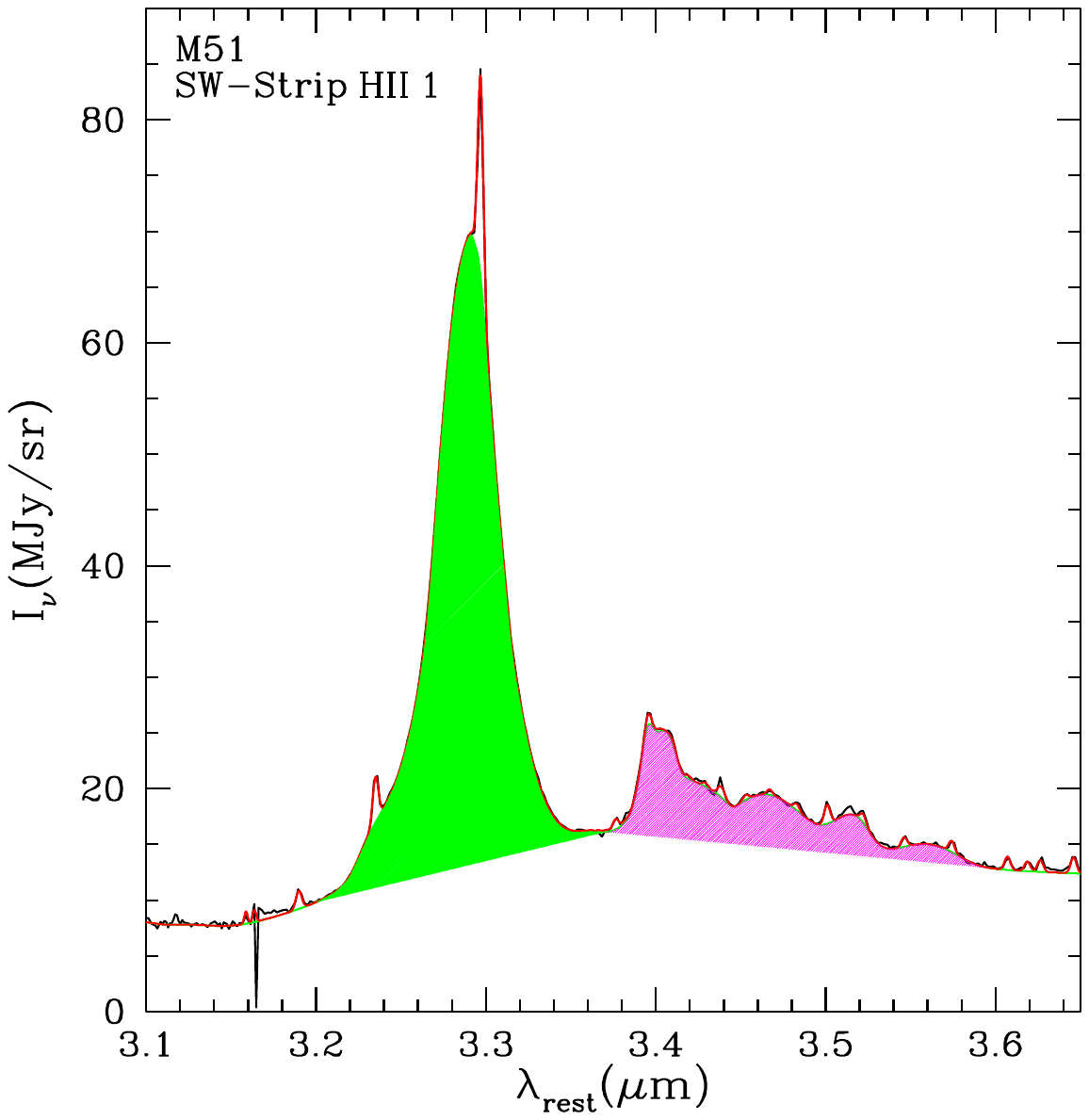}
\includegraphics[angle=0,width=\fwidth,
                 clip=true,trim=0.5cm 5.0cm 0.0cm 2.0cm]
{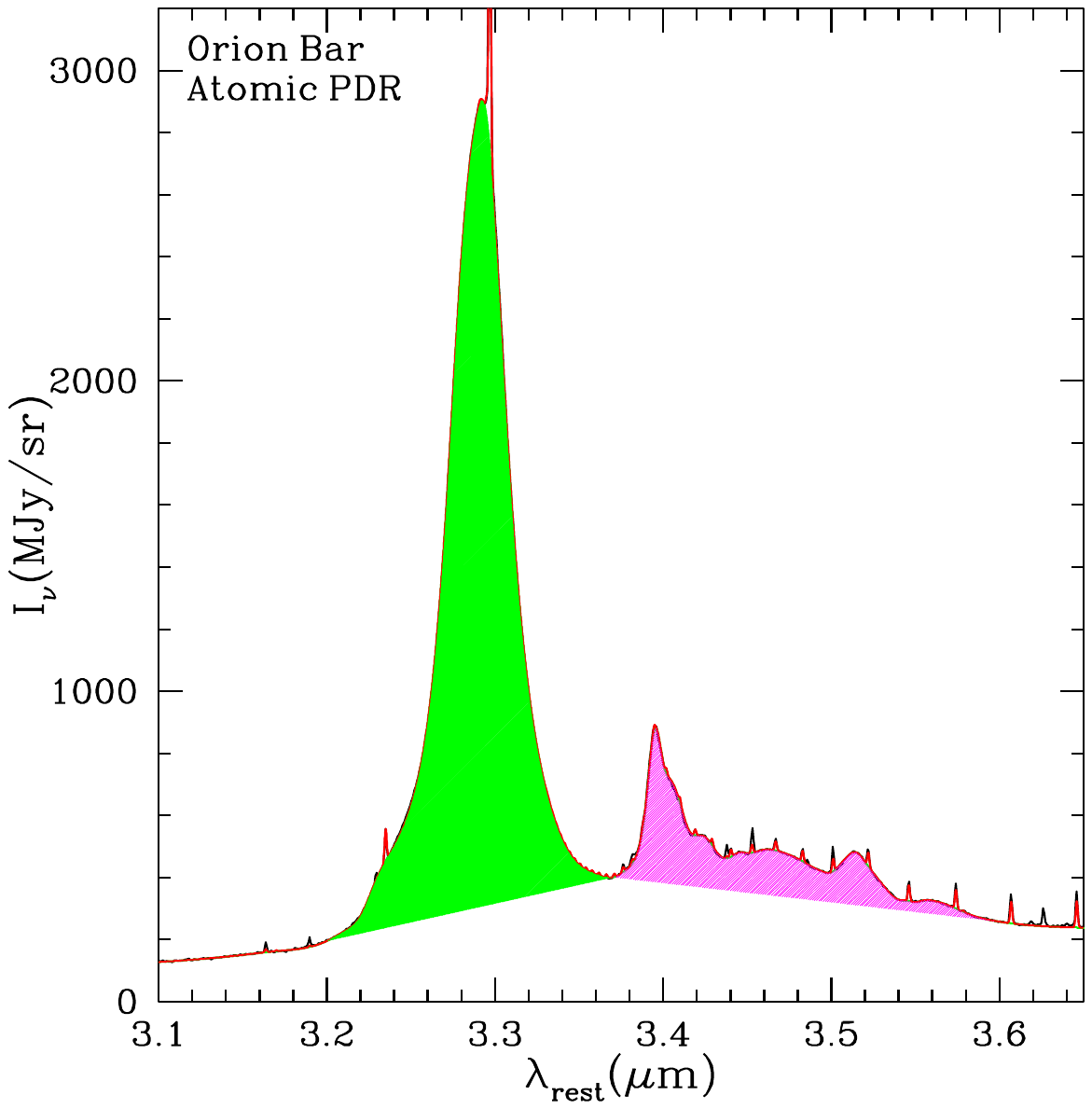}
\caption{\label{fig:CH}\footnotesize The $3-4\micron$ spectra of
  SW-Strip \ion{H}{2} in M51 (left) and the ``atomic PDR'' in the
  Orion Bar (right).  Red curve: model spectrum including emission
  lines.  Areas shaded green and magenta show how the integrated
  intensities $F_\aromCH^{\rm clip}$ and $F_\nonaromCH^{\rm clip}$ are
  evaluated (see text).}  
\end{center}
\end{figure}

Figure \ref{fig:33+34} shows normalized $3.1-3.65\micron$ emission
spectra (after removal of emission lines by interpolation) from the
four \ion{H}{2} region extractions in M51, the two Orion Bar
positions, and the two positions in M17.  In addition to the strong
peak at $3.29\micron$, all eight spectra exhibit a broad feature
extending from $3.37$ to $3.60\micron$.  This complex appears
equivalent to the broad ``plateau'' at $3.47\micron$ found at lower
resolution by \citet{Lai+Smith+Baba+etal_2020}, resolved at higher
resolution into peaks at $3.402$, $3.465$, $3.518$, and $3.56\micron$.

The $3.29\micron$ feature is the aromatic C--H stretching mode in
polycyclic aromatic hydrocarbons
\citep{Leger+Puget_1984,Allamandola+Tielens+Barker_1985,Tielens_2008}.
The similarity of the $3.29\micron$ feature in the eight spectra is
striking.  Except for the wings, the $3.29\micron$ feature can be
approximated by a Drude profile
\beq \label{eq:drude}
I_\nu^{\rm CD} = B
\frac{\gamma^2}
     {\left(\lambda/\lambda_0-\lambda_0/\lambda\right)^2+\gamma^2}
~~~.
\eeq
with central wavelength $\lambda_\aromCH=3.289\micron$ and
$\FWHM=0.0423\micron$ ($\gamma_\aromCH=0.013$), as shown in Figure
\ref{fig:33+34}.

The broad emission feature extending from $3.37-3.60\micron$
presumably includes a small contribution from aromatic C--H overtone
emission \citep{Allamandola+Tielens+Barker_1989}, but is dominated by
C--H stretching modes in nonaromatic material.  The 3.40$\micron$
feature [which \citet{Peeters+Habart+Berne+etal_2024} find to be
  composed of sub-components centered at $3.395$, $3.403$, and
  $3.424\micron$] is generally ascribed to the C--H stretch in
methylene ($-\Ca\Ha_2-$) groups in aliphatic hydrocarbons
\citep{Allamandola+Tielens+Barker_1989}.  The $3.46\micron$ and
$3.52\micron$ emission features have been attributed to H attached to
diamondlike carbon \citep{Allamandola+Sandford+Tielens+Herbst_1992,
  VanKerckhoven+Tielens+Waelkens_2002}.

Let $F_\aromCH$ be the power in the $3.29\micron$ aromatic C--H
stretch feature and $F_\nonaromCH$ be the power in the
$3.37-3.60\micron$ nonaromatic C--H stretch features.  Decomposition
of the observed emission into components plus an underlying continuum
is problematic because of both blending and the uncertain nature of
the underlying emission ``plateau''.\footnote{%
See, e.g., the alternative decompositions shown in Figure 16 of 
\citet{Peeters+Habart+Berne+etal_2024}.}

We quantify $F_\aromCH$ and $F_\nonaromCH$ using the model-free
empirical ``clipping'' method\footnote{
$\lambda F_\lambda^{\rm cut}=a\ln \lambda + b$ is defined passing
through $\lambda F_\lambda^{\rm obs}$ at ``clip points''
$\lambda_1$,$\lambda_2$ = $3.20,3.37\micron$ for the aromatic feature,
and $3.37,3.60\micron$ for the nonaromatic feature.  The power in each
feature is $F^{\rm clip}=\int_{\lambda_1}^{\lambda_2} (F_\lambda^{\rm
  obs}-F_\lambda^{\rm cut})d\lambda$.}
described in \citet{Draine+Li+Hensley+etal_2021}.  Figure \ref{fig:CH}
shows two examples of this clipping procedure; the shaded areas show
the integrated power included in $F_\aromCH^{\rm\,clip}$ and
$F_\nonaromCH^{\rm\,clip}$.  Correcting for extinction,
\beqa
F_\aromCH^{\rm\,clip,corr} &~=~& e^{\tau_{3.29}}F_\aromCH^{\rm\,clip}
\\
F_\nonaromCH^{\rm\,clip,corr} &=& e^{\tau_{3.40}}F_\nonaromCH^{\rm\,clip}
~~~,
\eeqa
where $\tau_{3.29}$ and $\tau_{3.40}$ are obtained from the
\citet{Gordon+Clayton+Decleir+etal_2023} extinction curve with $A_V$
determined from the \ion{H}{1} recombination lines.  Define
\beq \label{eq:fdef}
f\equiv \frac{F_\aliphCH}{F_\nonaromCH^{\rm clip}}
\eeq
to be the ratio of the true aliphatic C--H emission to the ``clipped''
measure of the nonaromatic C--H emission.  Blending and uncertainties
about the level of the emission underlying the nonaromatic features
make it difficult to determine $f$; we take $f\approx 1.0$ as a
working estimate.

\subsection{\label{sec:aliphCD} Emission from Aliphatic C--D}

\begin{figure}
\begin{center}
\includegraphics[angle=0,width=\fwidth,
               clip=true,trim=0.5cm 5.cm 0.0cm 4.5cm]
{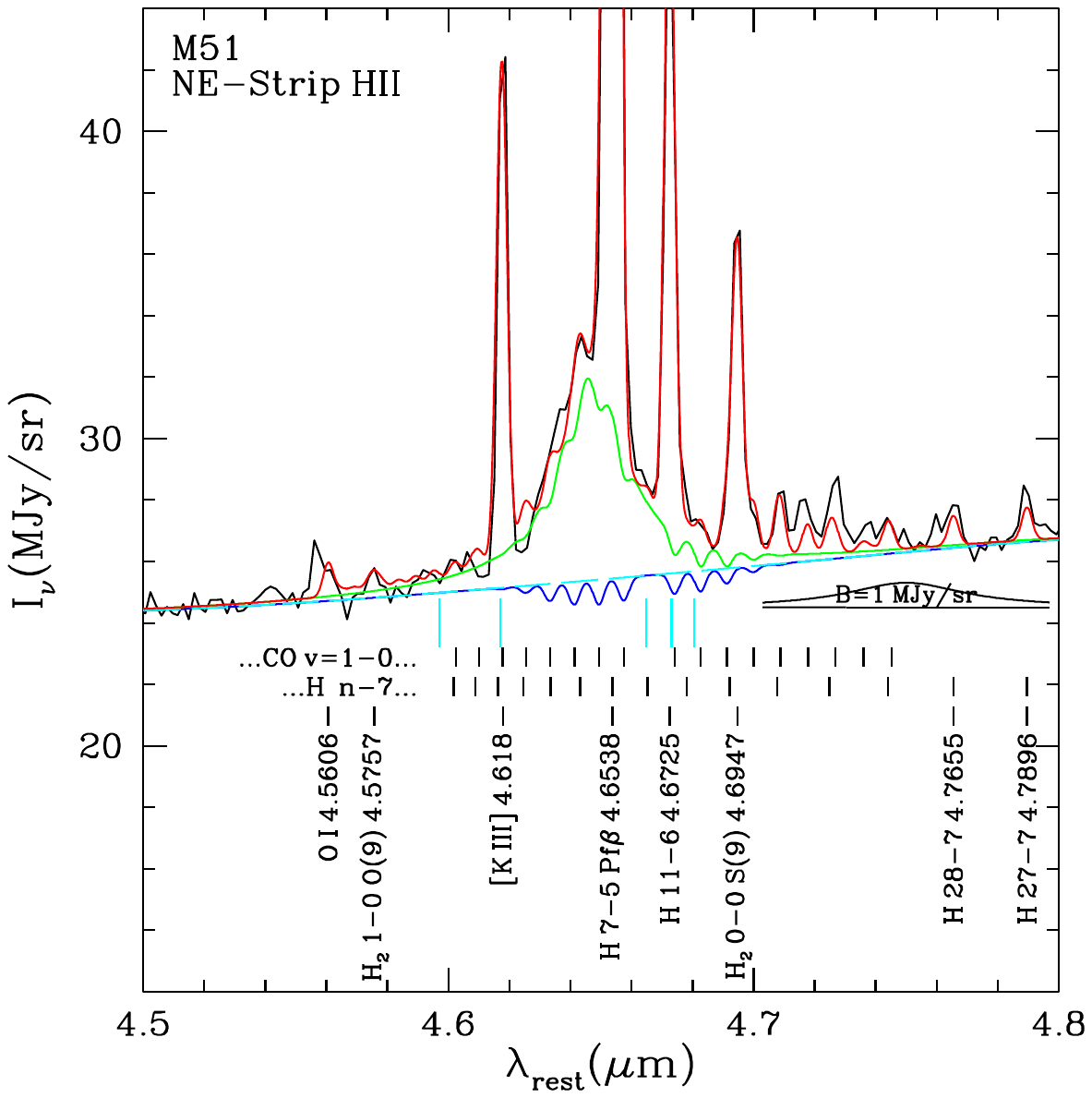}
\includegraphics[angle=0,width=\fwidth,
               clip=true,trim=0.5cm 5.cm 0.0cm 4.5cm]%
{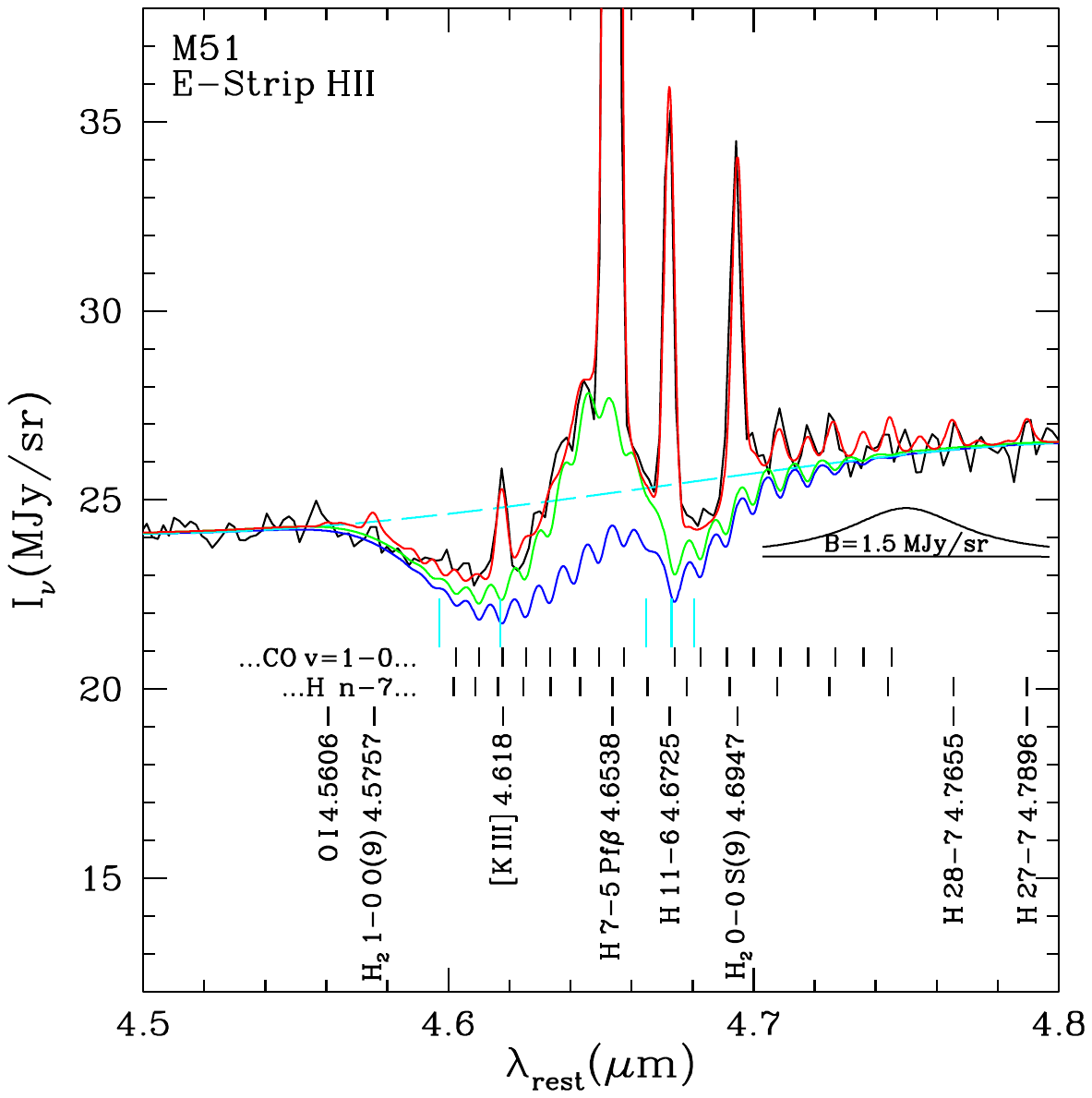}
\includegraphics[angle=0,width=\fwidth,
               clip=true,trim=0.5cm 5.cm 0.0cm 4.5cm]%
{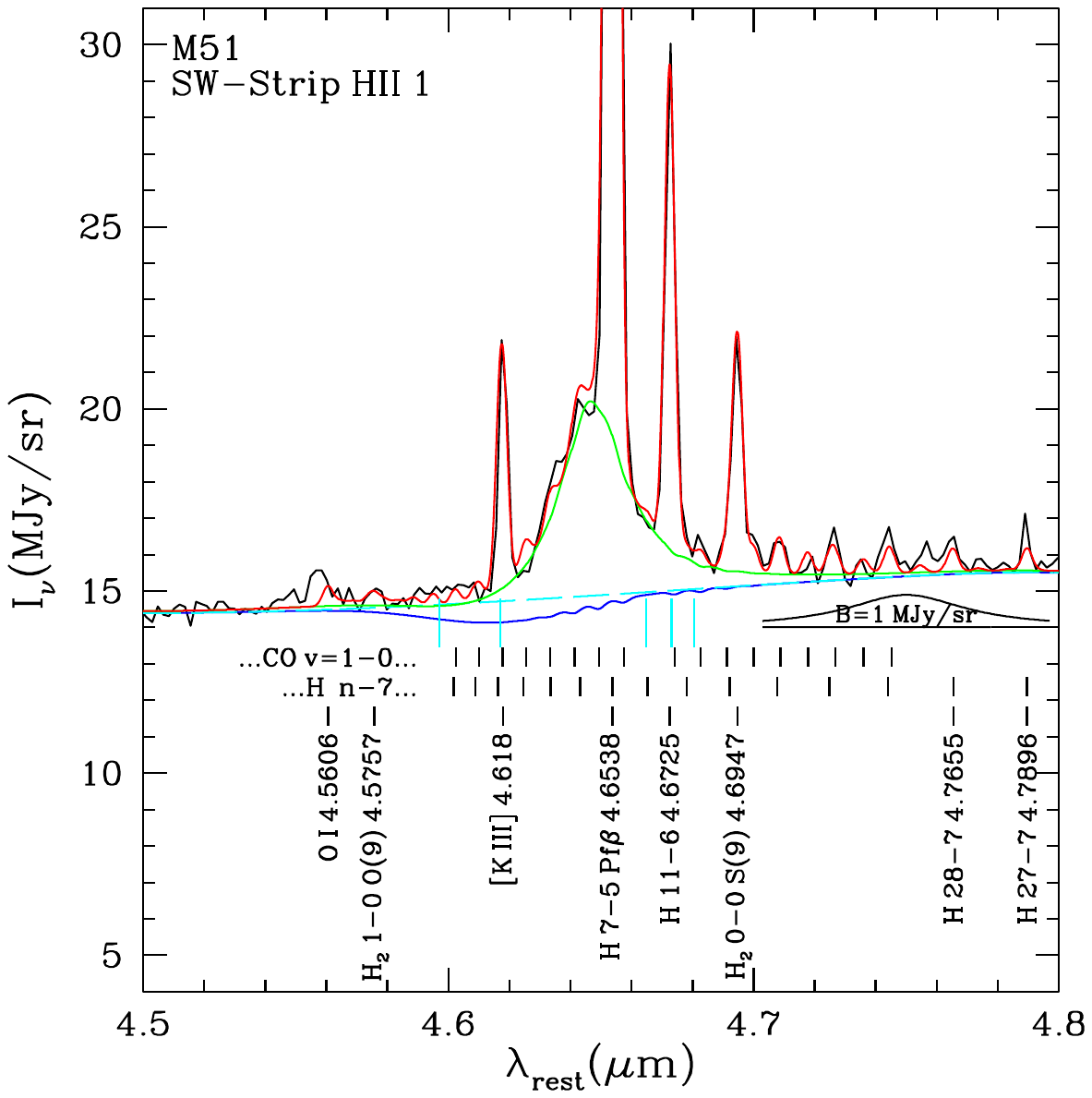}
\includegraphics[angle=0,width=\fwidth,
               clip=true,trim=0.5cm 5.cm 0.0cm 4.5cm]%
{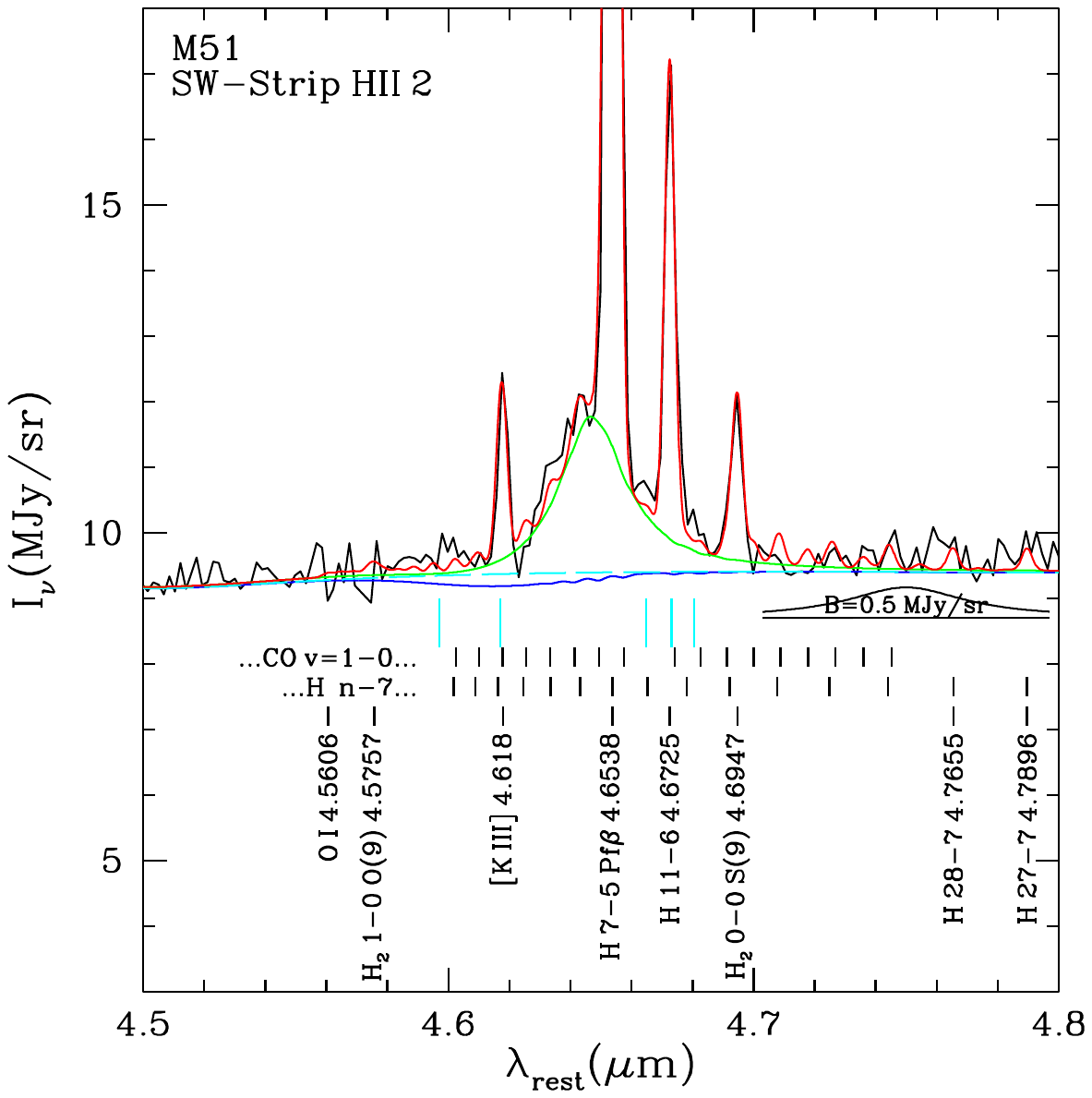}
\caption{\label{fig:m51aliphCD}\footnotesize The $4.5-4.8\micron$
  spectra of the four sightlines in Figure \ref{fig:m51_4-5}.  Black
  curves: observed spectra.  Dashed cyan curves: estimate for the
  dust-attenuated ``continuum'' (not including $4.65\micron$ feature),
  but without absorption by ice or CO. Solid blue curves: continuum
  after absorption by ice and $v=0$ CO.  Green curves: continuum plus
  4.65$\micron$ emission feature (Eq.\ \ref{eq:drude}) with
  attenuation by dust, ice, and CO.  Red curves: green curves plus
  emission lines of H, He, H$_2$, $v=1$\,CO, and [\ion{K}{3}] (see
  text). Inset profile: upper limit on an emission feature at
  $4.75\micron$ (see text).
  }
\end{center}
\end{figure}
\begin{figure}
\begin{center}
\includegraphics[angle=0,width=\fwidth,
               clip=true,trim=0.5cm 5.cm 0.0cm 4.5cm]%
{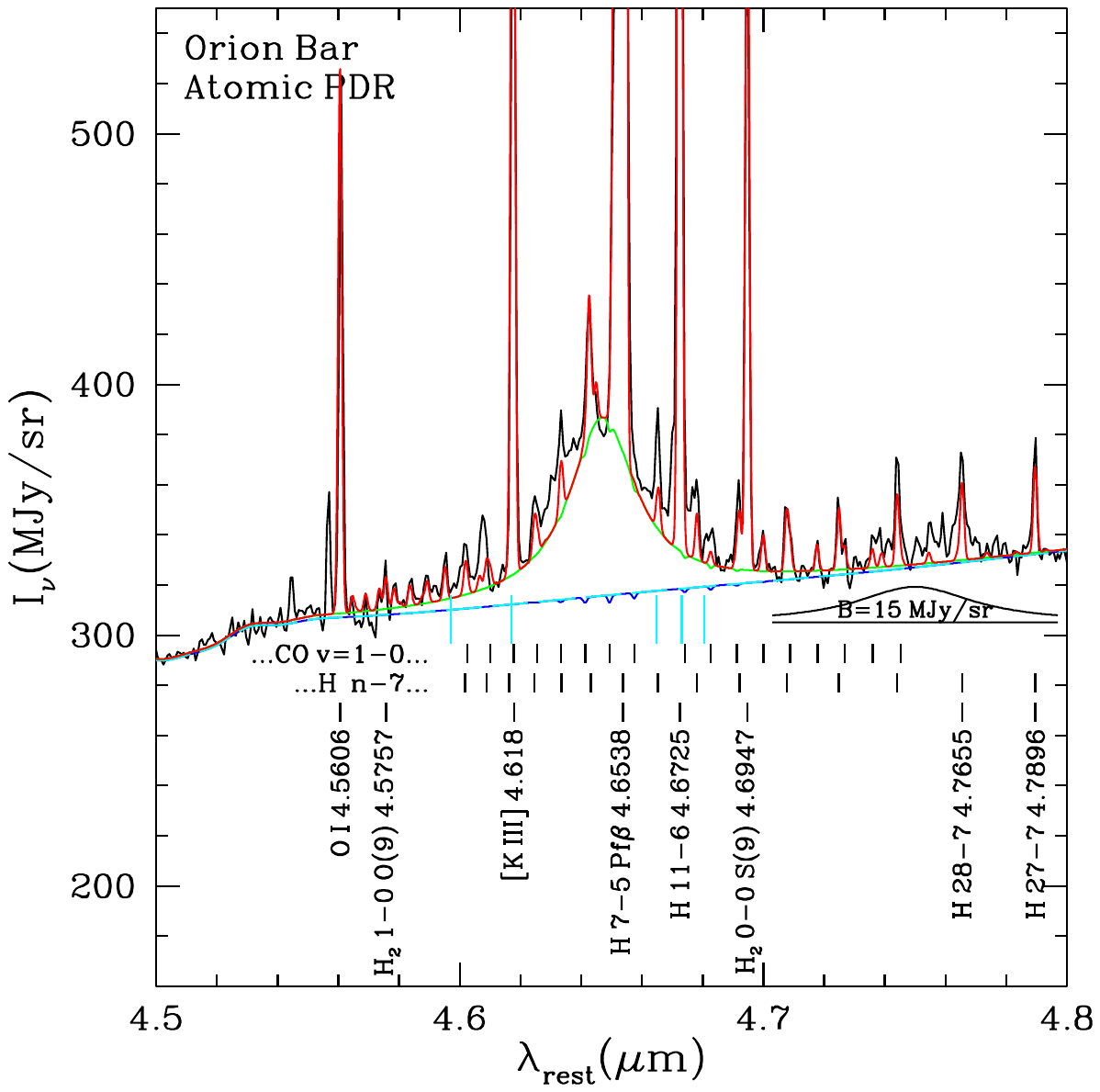}
\includegraphics[angle=0,width=\fwidth,
               clip=true,trim=0.5cm 5.cm 0.0cm 4.5cm]%
{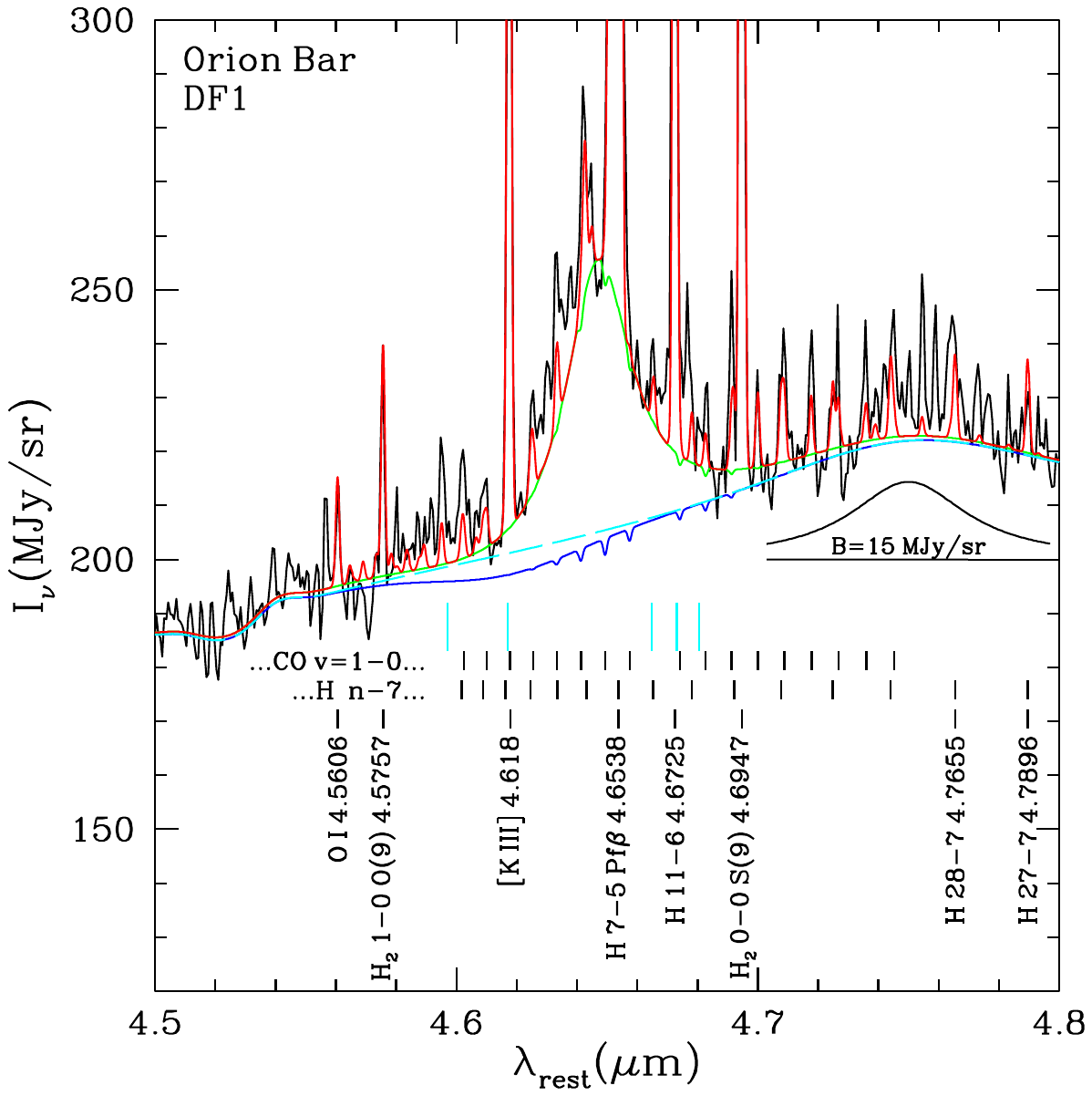}
\includegraphics[angle=0,width=\fwidth,
               clip=true,trim=0.5cm 5.cm 0.0cm 4.5cm]%
{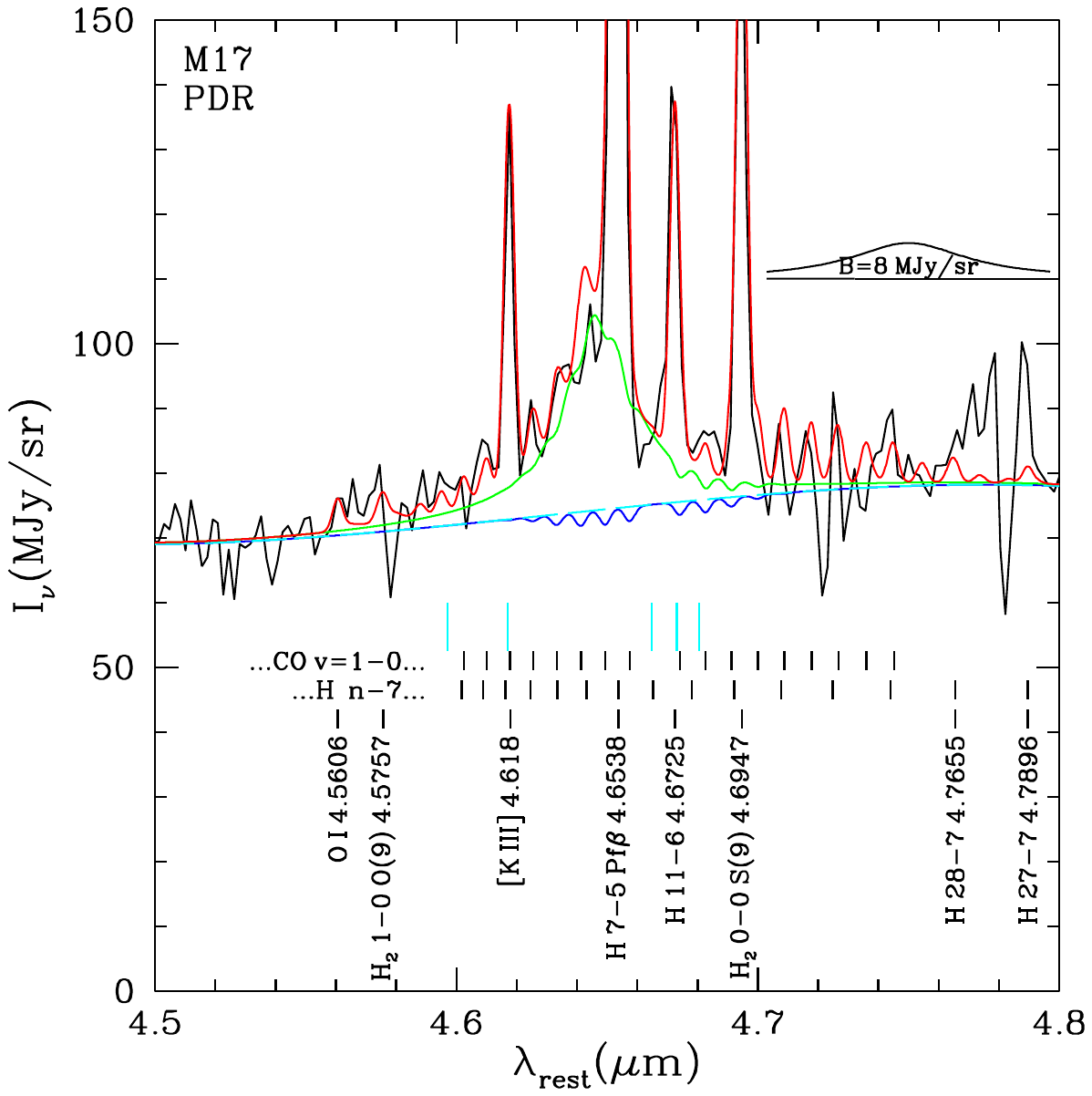}
\includegraphics[angle=0,width=\fwidth,
               clip=true,trim=0.5cm 5.cm 0.0cm 4.5cm]%
{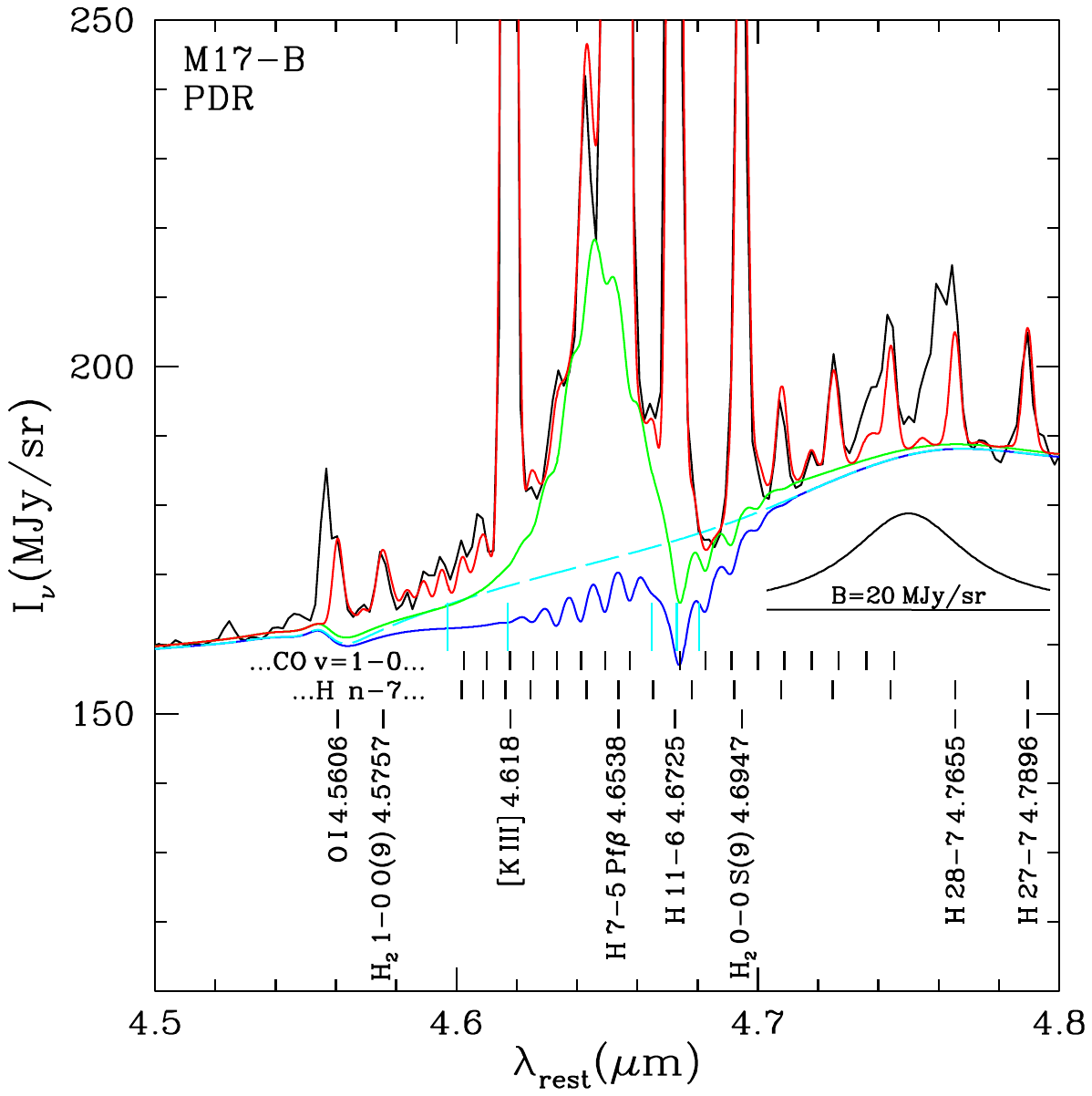}
\caption{\label{fig:orion+m17aliphCD}\footnotesize Same as Figure
  \ref{fig:m51aliphCD}, but for two pointings in the Orion Bar
  \citep{Peeters+Habart+Berne+etal_2024}, and two positions in the M17
  PDR \citep{Boersma+Allamandola+Esposito+etal_2023} (see text).}
\end{center}
\end{figure}

Figure \ref{fig:m51aliphCD} shows $4.5-4.8\micron$ spectra for the
four regions in M51; Figure \ref{fig:orion+m17aliphCD} shows the same
spectral region for two positions in the Orion Bar
\citep{Peeters+Habart+Berne+etal_2024} and two positions in the M17
PDR \citep{Boersma+Allamandola+Esposito+etal_2023}.  As evident in
Figure \ref{fig:m51_4-5}, the M51 spectra include a significant dust
emission feature centered near $4.65\micron$, near the expected
wavelength of the aliphatic C--D stretch, similar to the $4.65\micron$
feature found in M17 by \citet{Boersma+Allamandola+Esposito+etal_2023}
and in the Orion Bar by \citet{Peeters+Habart+Berne+etal_2024}.

To fit this feature, for all regions we adopt a simple Drude profile
(Equation (\ref{eq:drude})) with central wavelength
$\lambda_\aliphCD=4.647\micron$ and $\FWHM=0.0265\micron$
($\gamma_\aliphCD=0.0057$).  The peak intensity $B_\aliphCD$ is
adjusted so that Equation (\ref{eq:I_nu}) approximates the observed
emission for each region, as shown in Figures \ref{fig:m51aliphCD} and
\ref{fig:orion+m17aliphCD}.  There is no reason to expect the actual
emission profile to conform to Equation (\ref{eq:drude}), but with the
signal-to-noise ratio of the present spectra, in a region crowded with
numerous emission lines, our Drude profile provides an acceptable fit.
Future observations may permit refinement of the aliphatic C--D
emission profile.

For the Drude profile, the extinction-corrected integrated intensity
\beq
F_\aliphCD^{\rm corr} = 
\int I_\nu^{\rm CD} \, d\nu 
=
\frac{\pi c}{2}\frac{\gamma_\aliphCD B_\aliphCD}{\lambda_\aliphCD}
\eeq
for each region is given in Table \ref{tab:params}.

The aliphatic C-D stretch was previously detected in M17-B-PDR
\citep{Boersma+Allamandola+Esposito+etal_2023} and in the Orion Bar
atomic PDR and DF1 \citep{Peeters+Habart+Berne+etal_2024}.  Table
\ref{tab:compare} compares those results with our estimates for these
features.  We also compare our upper limit for the 4.75$\micron$
feature in the Orion Bar with the feature intensity measured by
\citet{Peeters+Habart+Berne+etal_2024}.

\begin{table}
\begin{center}
\caption{\label{tab:compare}Comparison with Previous Work}
{\footnotesize
\begin{tabular}{l l c c l}
\hline
Feature            & Region       & This Work$^a$ 
                   & Previous Work & Reference \\ 
                   & & $(10^{-6}\erg\cm^{-2}\s^{-1}\sr^{-1})$
                   & $(10^{-6}\erg\cm^{-2}\s^{-1}\sr^{-1})$
\\
\hline
Aliphatic CD       & Orion Bar, atomic PDR  
                   & $410\pm60$ 
                   & $350$ 
                   &\citep{Peeters+Habart+Berne+etal_2024}
\\
''                 & Orion Bar, DF1   
                   & $300\pm45$ & $220$
                   &\citep{Peeters+Habart+Berne+etal_2024}
\\
''                 & M17-B-PDR    
                   & $280\pm60$ & $273\pm52$
                   & \citep{Boersma+Allamandola+Esposito+etal_2023}
\\
\hline
$4.75\micron$ feature
                   & Orion Bar, atomic PDR
                   & $<140$     & $50$
                   & \citep{Peeters+Habart+Berne+etal_2024}
\\
''                 & Orion Bar, DF1
                   & $<140$     & $90$
                   & \citep{Peeters+Habart+Berne+etal_2024}
\\
\hline
\multicolumn{5}{l}{$a$ Not corrected for reddening.}
\end{tabular}
} 
\end{center}
\end{table}

\subsection{Regional Variations in Deuteration}

\begin{figure}
\begin{center}
\includegraphics[angle=0,width=\fwidth,
                 clip=true,trim=0.5cm 5.0cm 0.0cm 2.0cm]
{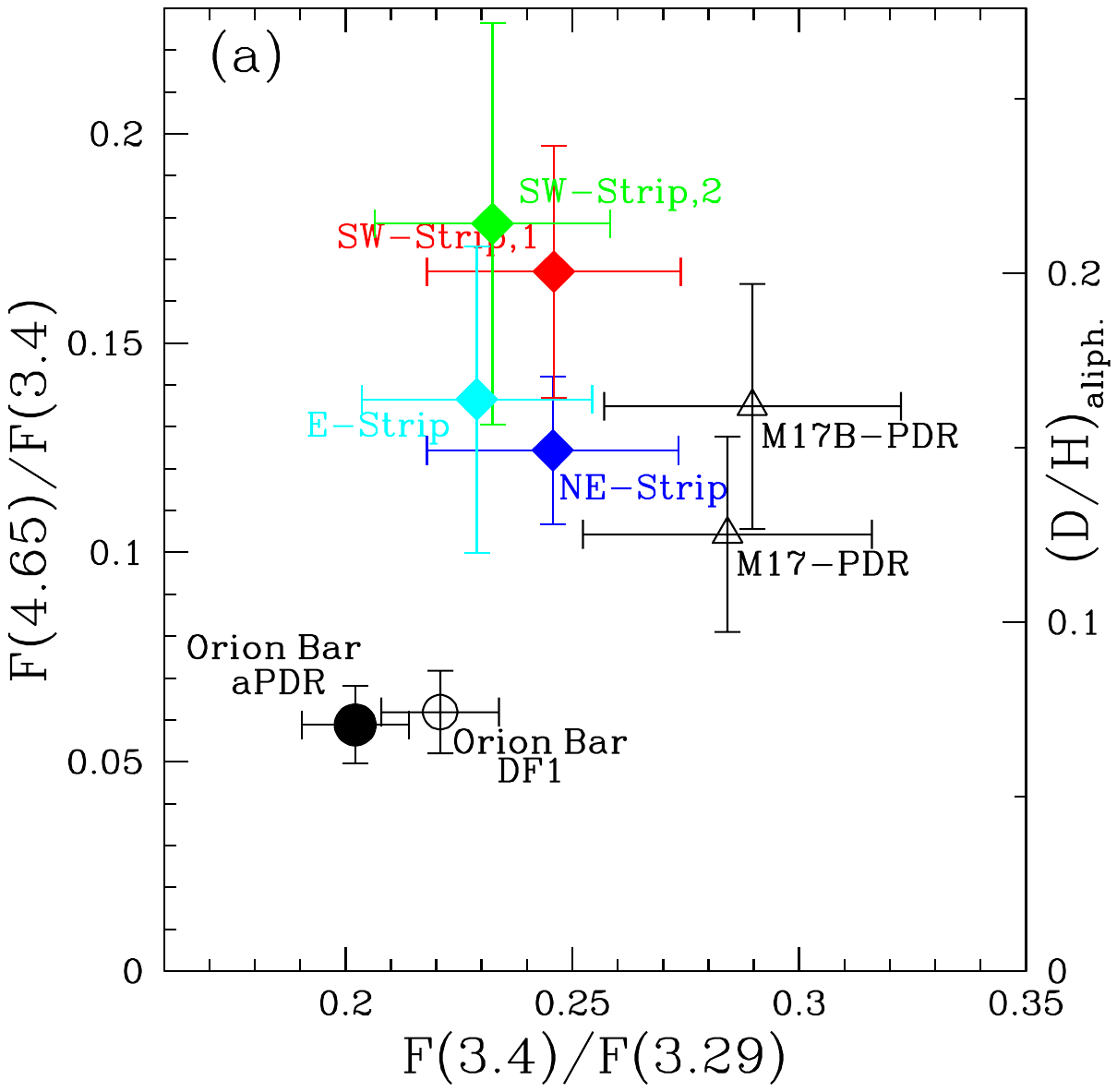}
\includegraphics[angle=0,width=\fwidth,
                 clip=true,trim=0.5cm 5.0cm 0.0cm 2.0cm]
{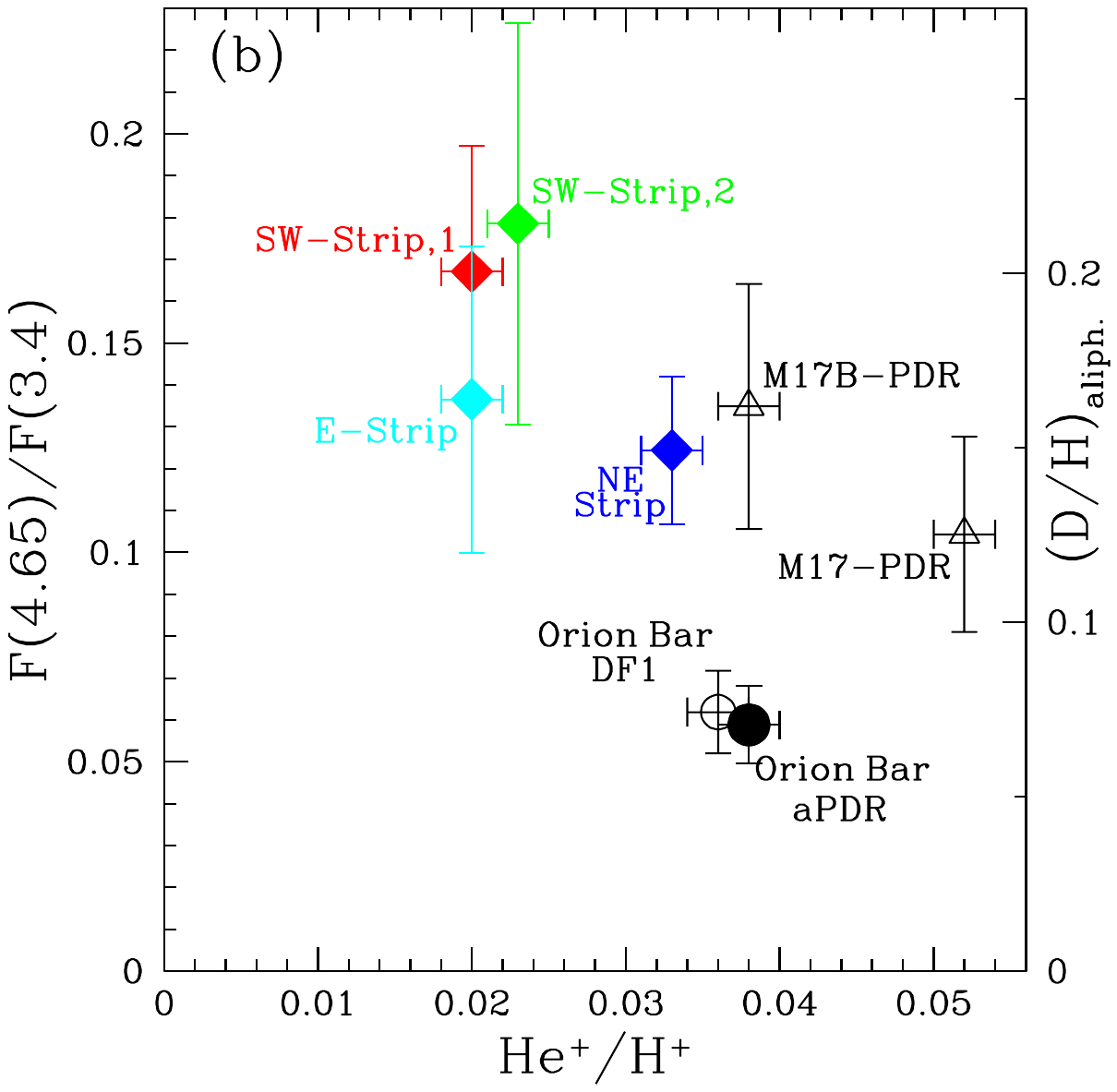}
\caption{\label{fig:cdr}\footnotesize (a) $F(4.65)/F(3.4)$
  vs.\ $F(3.4)/F(3.29)$, where $F(4.65)$ is the power in the
  $4.65\micron$ aliphatic C-D feature, $F(3.4)$ is the ``clipped''
  estimate of the non-aromatic C-H stretch power, and $F(3.29)$ is the
  ``clipped'' estimate of the aromatic C-H stretch power (see Figure
  \ref{fig:CH}).  The right-hand scale shows the estimated
  $(\Da/\Ha)_\aliph$ in the emitting nanoparticles, assuming
  $K/f=1.2$.  Deuteration is 2-3 times higher in M51 than in the Orion
  Bar.  $(\Da/\Ha)_\aliph$ does not appear to be correlated with
  nonaromatic fraction.  (b) Same as (a), but plotted against
  $\He^+/\Ha^+$.  $(\Da/\Ha)_\aliph$ appears to depend on the hardness
  of the radiation ionizing the \ion{H}{2} region associated with the
  PDR where the PAHs are located.  }
\end{center}
\end{figure}

The strength of the $4.65\micron$ aliphatic C--D emission relative to
the $3.37-3.60\micron$ nonaromatic C--H emission varies significantly
among the four regions in M51 -- see Figure \ref{fig:cdr}a, which
plots $F(4.65\micron)/F(3.4\micron)$ for the four regions in M51, and
also for two positions in the Orion Bar, and two positions in the M17
PDR.  What is responsible for the variations in
$F(4.65\micron)/F(3.4\micron)$?

Previous studies have shown that the aliphatic/aromatic ratio shows
regional variations in the Milky Way and other galaxies
\citep{Lai+Smith+Baba+etal_2020,Yang+Li_2023b}.  However, in Figure
\ref{fig:cdr}a there is no apparent correlation of
$F(4.65\micron)/F(3.4\micron)$ with $F(3.4\micron)/F(3.29\micron)$.
Evidently, the environmental effects responsible for the variations in
the aromatic/nonaromatic ratio do not directly lead to variations in
deuteration.

Another quantity that varies from region to region is $\He^+/\Ha^+$,
measuring the degree of helium ionization in associated \ion{H}{2}
regions.  Figure \ref{fig:cdr}(b) plots $F(4.65\micron)/F(3.4\micron)$
versus $\He^+/\Ha^+$.  At least in this sample,
$F(4.65\micron)/F(3.4\micron)$ shows a negative correlation with
$\He^+/\Ha^+$.  Possible explanations for this trend are discussed
below in Section \ref{sec:interp}.

\subsection{The D/H Ratio in the Nanoparticles}

The integrated absorption cross section (``band strength'')
$A\equiv\int\sigma_\nu d\lambda^{-1}$ per C--D or C--H bond is expected
to be smaller for C--D (relative to C--H) by about a factor 
$A_\CH/A_\CD\approx1.75$
\citep{Bauschlicher+Langhoff+Sandford+Hudgins_1997}.\footnote{%
\citet{Mori+Onaka+Sakon+etal_2022} find $A_\CH/A_\CD=2.6$ for the
aliphatic bonds.  This would raise the inferred D/H ratio by a factor
$\sim$$1.5$.}
The D/H ratio of the aliphatic material in the emitting nanoparticles
is related to the observed emission by
\beq
\left(\frac{\Da}{\Ha}\right)_{\rm aliph.}  
= K \times
\frac{F_\aliphCD^{\rm\,corr}}{F_\aliphCH^{\rm\,corr}} 
=
\frac{K}{f} \times
\frac{F_\aliphCD^{\rm\,corr}}{F_\nonaromCH^{\rm\,clip,corr}}
~~~, 
\eeq
where $f\approx1$ is the factor relating the aliphatic C--H emission
to $F_\nonaromCH^{\rm\,clip,corr}$ (Equation \ref{eq:fdef}), and
\beq \label{eq:K}
K = 1.75
  \left(\frac{\lambda_\CD}{\lambda_\CH}\right)^3 \left(\frac{\int
    [e^{hc/\lambda_\CH\kB T}-1]^{-1} (dP/dT)~ dT} {\int
    [e^{hc/\lambda_\CD\kB T}-1]^{-1} (dP/dT)~ dT} \right)
 ~~~.  
\eeq
Here $dP/dT$ is the temperature distribution function for the
nanoparticles responsible for the $\lambda < 5\micron$ emission.
Estimation of D/H requires knowledge of the temperature distribution
function $dP/dT$, requiring modeling of the stochastic heating process
\citep{Draine+Li_2001}, which depends on the particle size.  However,
if we assume that the emitting material can be characterized by a
temperature in $500\K <T<1500\K$, then we find $0.5 < K < 1.9$, i.e.,
\beq
K \approx 1.2\pm0.7
~~~,
\eeq
%
%
%
with $K=1.2$ corresponding to an emitting temperature $T=900\K$.
In the discussion below, we take $K=1.2$ and $f=1$.


\begin{table}
\newcommand \fna {a}
\newcommand \fnb {b}
\begin{center}
\caption{\label{tab:D/H} Deuteration in Modeled Regions}
{\footnotesize
\begin{tabular}{c c c}
\hline
region&  (D/H)$_{\rm aliphatic}$$^\fna$ & (D/H)$_{\rm aromatic}$$^\fnb$ \\
\hline
M51:~NE-Strip        & $0.15\pm 0.02$ & $<0.007$\\
M51:~SW-Strip,1      & $0.20\pm 0.04$ & $<0.006$\\
M51:~SW-Strip,2      & $0.21\pm 0.06$ & $<0.016$\\
M51:~E-Strip         & $0.17\pm 0.04$ & $<0.013$\\
Orion Bar:~atomic PDR& $0.071\pm0.011$ & $<0.003$\\
Orion Bar:~DF1       & $0.074\pm0.012$ & $<0.005$\\
M17-PDR              & $0.125\pm0.028$ & $<0.022$\\
M17-B-PDR            & $0.16\pm0.04$   & $<0.018$\\
\hline
\multicolumn{3}{l}{$\fna$ Assuming $K=1.2$ and
                   $F_\aliphCH/F_\nonaromCH^{\rm clip} = 1$ (see text).}\\
\multicolumn{3}{l}{$\fnb$ Assuming $K=1.2$ and 
                   $F_\aromCH/F_\aromCH^{\rm clip}=1/0.71$ (see text).}\\
\end{tabular}}
\end{center}
\end{table}


Results for $(\Da/\Ha)_\aliph$ are given in Table \ref{tab:D/H}, and
on the right-hand scale in Figure \ref{fig:cdr}b.  The four positions
in M51 show $(\Da/\Ha)_{\rm aliph}$ varying by a factor of $\sim$1.3,
from $0.15$ in NE-Strip to $0.21$ in SW-Strip,2.  The four positions
in M51 have
%
\beq
\langle\Da/\Ha\rangle_{\rm aliph} \approx 0.17 \pm 0.02
~~~.
\eeq
For the Orion Bar positions we find $\langle\Da/\Ha\rangle_{\rm
  aliph}= 0.07\pm0.01$, smaller than the values $0.23$ and $0.16$
found by \citet{Peeters+Habart+Berne+etal_2024}.  For M17 we find
$\langle\Da/\Ha\rangle_{\rm aliph}=0.14\pm0.02$, smaller than the
value $0.31\pm0.13$ found by
\citet{Boersma+Allamandola+Esposito+etal_2023}, and intermediate
between the Orion Bar and the four regions in M51.

The aliphatic material in M51 is D-enriched by a factor $\sim$$10^4$
relative to the overall abundance $\Da/\Ha\approx 23\ppm$.  It must be
stressed that $(\Da/\Ha)_\aliph$ is only determined for the
nanoparticles small enough ($N_\Ca \ltsim 10^3$ C atoms) that
single-photon heating (with $h\nu<13.6\eV$) can heat them to the high
temperatures ($T\gtsim500\K$) required to radiate effectively at
$4.6\micron$.  The majority of the hydrocarbon material is in larger
particles, which may have a different D/H ratio but are unconstrained
by $\lambda < 5\micron$ emission spectra.

The 4.65$\micron$ feature is narrow, with a fractional width
$\gamma_\CD \equiv \FWHM/\lambda_\CD \approx 0.006$ -- similar to the
fractional width of the $3.40\micron$ aliphatic component of the
nonaromatic C--H emission extending from $3.36-3.60\micron$ (see
Figure \ref{fig:33+34}).

\subsection{\label{sec:interp} Possible Interpretation of the $(\Da/\Ha)_\aliph$
-- $\He^+/\Ha^+$ Anticorrelation}

Figure \ref{fig:cdr}(b) shows that the highest degree of deuteration
is seen for the regions with the lowest values of $\He^+/\Ha^+$.  The
PAH emission in star-forming regions originates primarily in PDRs
adjacent to \ion{H}{2} regions.  As the stellar population ages, the
intensity of the $h\nu<13.6\eV$ radiation powering the PDRs declines,
and the far-UV spectrum softens \citep{Smith+Norris+Crowther_2002}.

One possible interpretation for the anticorrelation of aliphatic
deuteration and $\He^+/\Ha^+$ is that the $h\nu<13.6\eV$ radiation in
the PDRs is harsher -- both more intense, and with a harder spectrum
-- in regions with younger stars, as measured by $\He^+/\Ha^+$.  In
the quiescent ISM, single-photon heating events that lead to H loss
from small PAHs are relatively rare, and the small difference in
binding energy may lead to strong preferential retention of D, leading
to the observed high levels of deuteration.  However, in a high
intensity PDR with a hard FUV spectrum, the combination of more
frequent ``multiphoton'' heating events (where a photoabsorption
occurs before the PAH has fully cooled following the previous
photoabsorption) \emph{and} harder FUV photons may lead to rapid loss
of both D and H, and a drop in the degree of deuteration as the more
abundant H atoms from the gas replace the losses.  If the hydrocarbon
nanoparticles are deuterated prior to the arrival of the
photodissociation front and then lose deuterium in the PDR, perhaps
they lose deuterium more rapidly in the harsher PDRs near the youngest
stellar clusters.

Because loss of deuterium begins as soon as the nanoparticles are
exposed to FUV in the PDR, the particles responsible for emission in
the $4.65\micron$ C-D and $3.42\micron$ C-H features have already
undergone partial D loss -- the D/H ratio in the not-yet-exposed
nanoparticles should be higher than the value estimated from the
observed emission.

\subsection{Upper Limits on Aromatic C--D}

\begin{figure}
\begin{center}
\includegraphics[angle=0,width=\fwidth,
               clip=true,trim=0.5cm 5.cm 0.0cm 4.5cm]
{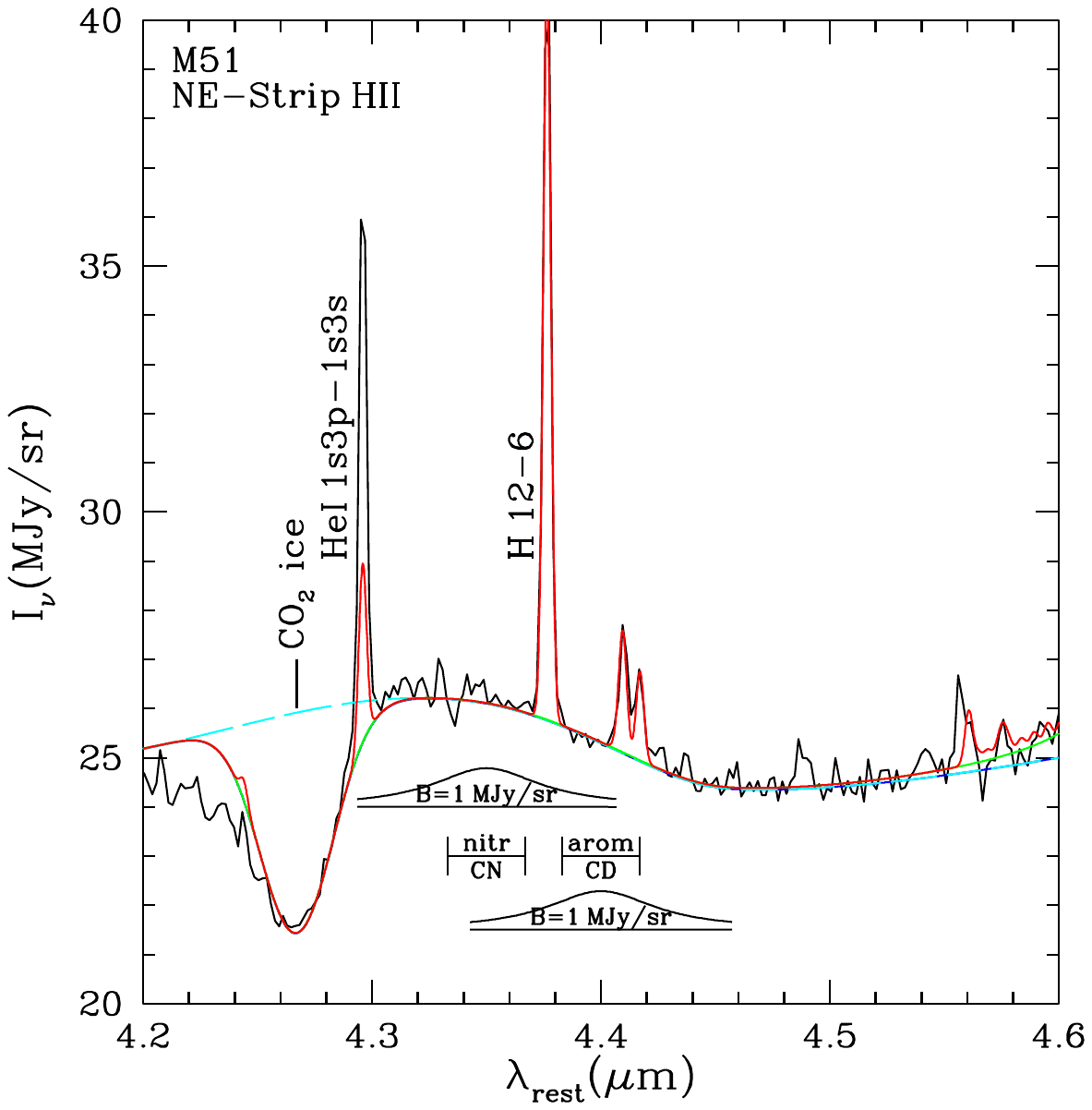}
\includegraphics[angle=0,width=\fwidth,
               clip=true,trim=0.5cm 5.cm 0.0cm 4.5cm]%
{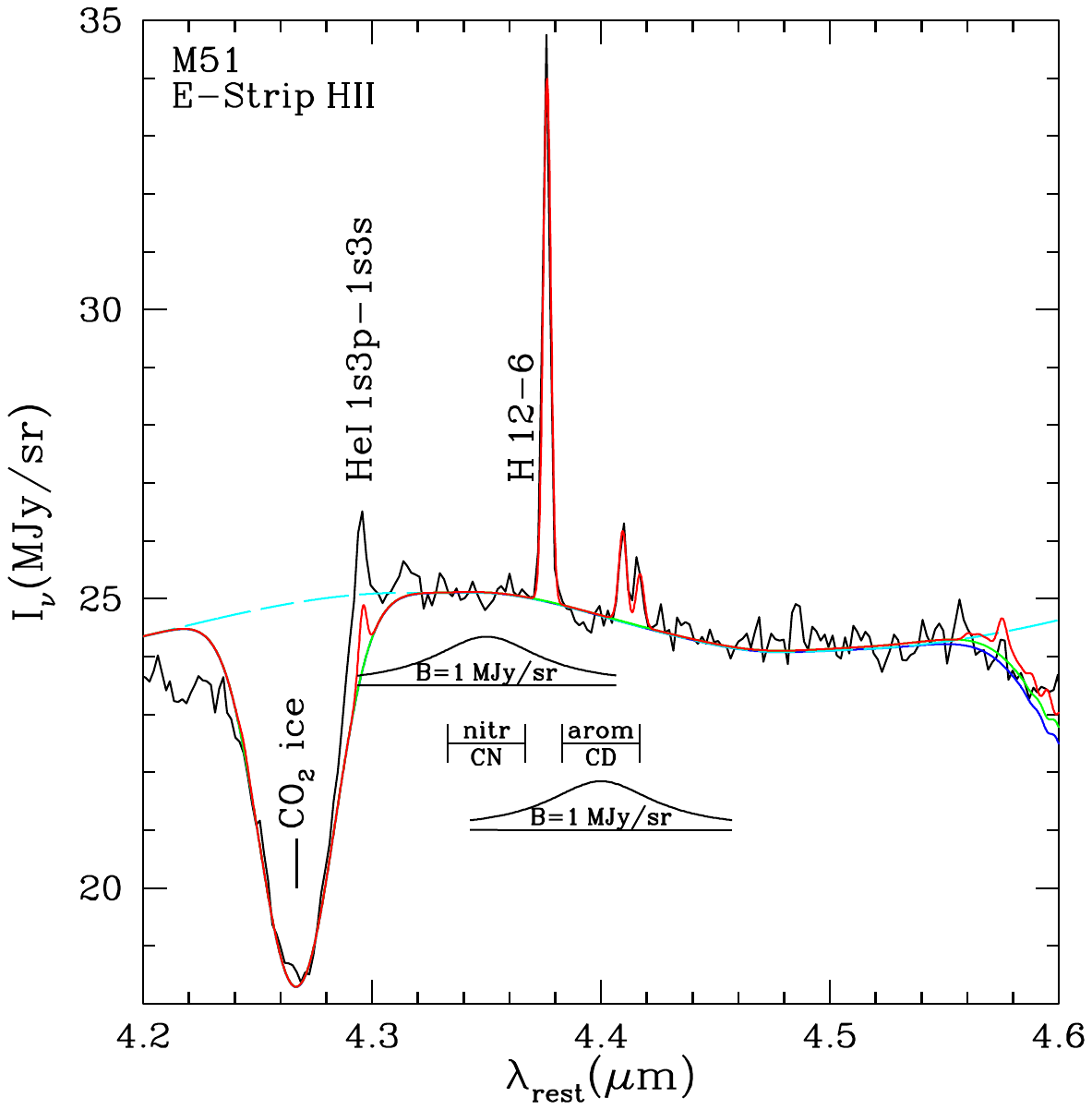}
\includegraphics[angle=0,width=\fwidth,
               clip=true,trim=0.5cm 5.cm 0.0cm 4.5cm]%
{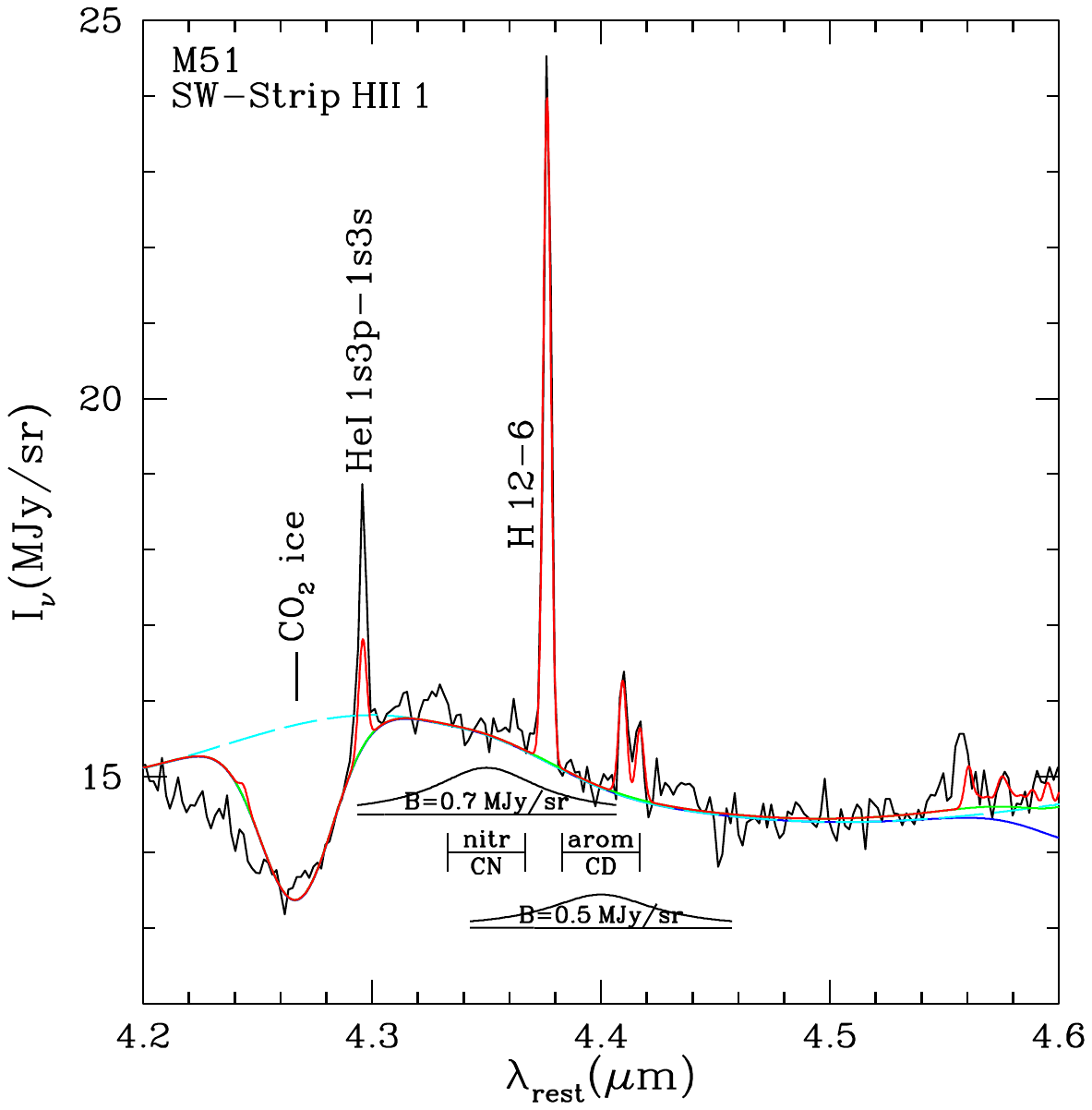}
\includegraphics[angle=0,width=\fwidth,
               clip=true,trim=0.5cm 5.cm 0.0cm 4.5cm]%
{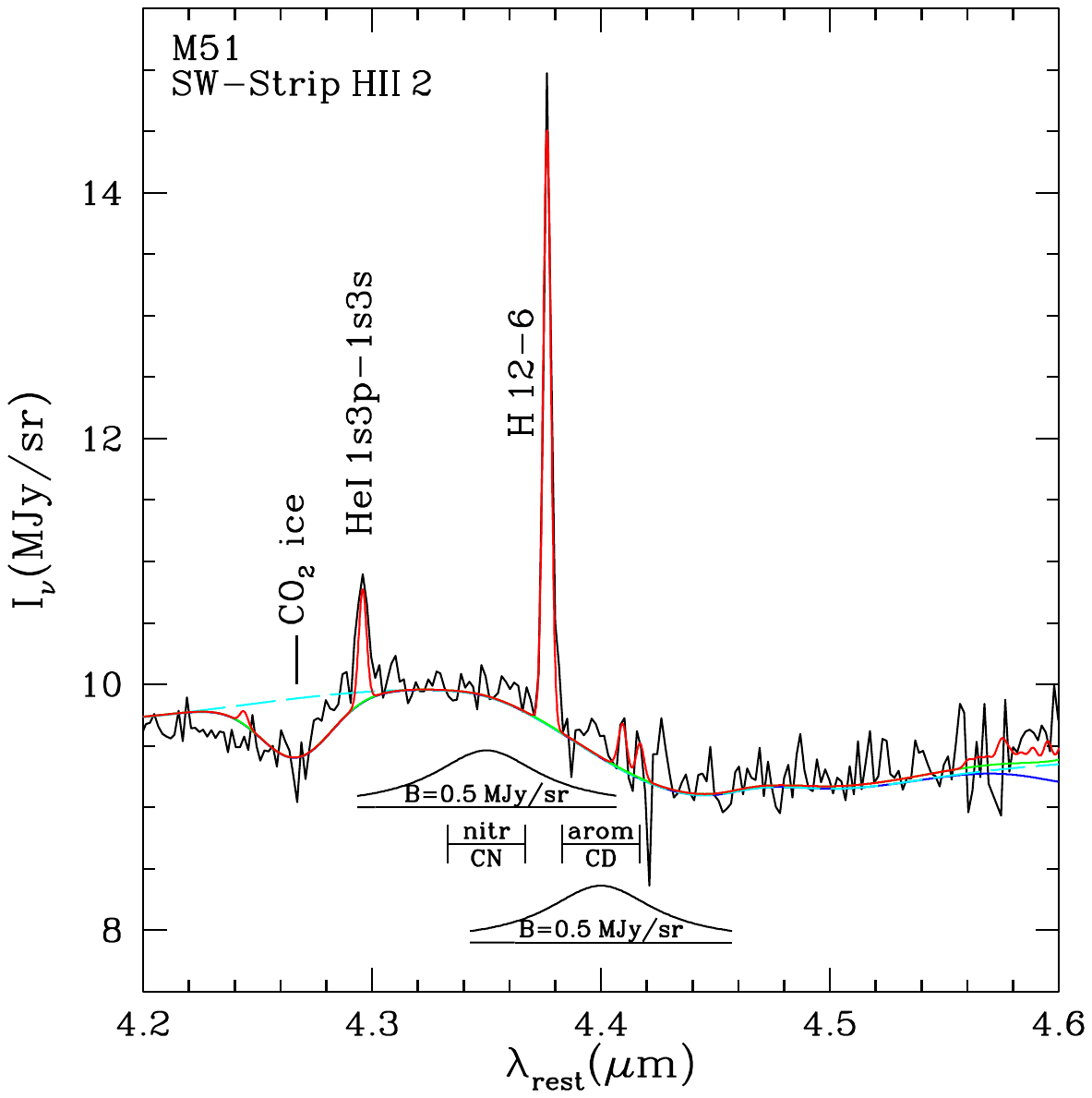}

\caption{\label{fig:m51aromCD}\footnotesize The $4.2-4.6\micron$
  spectra of the four sightlines in Figure \ref{fig:m51_4-5}.  Black
  curves: observed spectra.  Cyan dashed curves: estimate of continuum
  without ice absorption.  Green curves: continuum with ice absorption
  applied.  Red curves: continuum plus emission lines of H, He, and
  H$_2$ (see text).  The expected wavelengths for nitrile CN
  ($\sim$4.35$\mu$m) and aromatic C--D ($\sim$4.40$\mu$m) are
  indicated.  In the M51 spectra, there is no evidence of nitrile CN
  emission or aromatic C--D emission; the inset profile shows our
  estimated upper limit on the strength of either feature.}
\end{center}
\end{figure}
\begin{figure}
\begin{center}
\includegraphics[angle=0,width=\fwidth,
               clip=true,trim=0.5cm 5.cm 0.0cm 4.5cm]%
{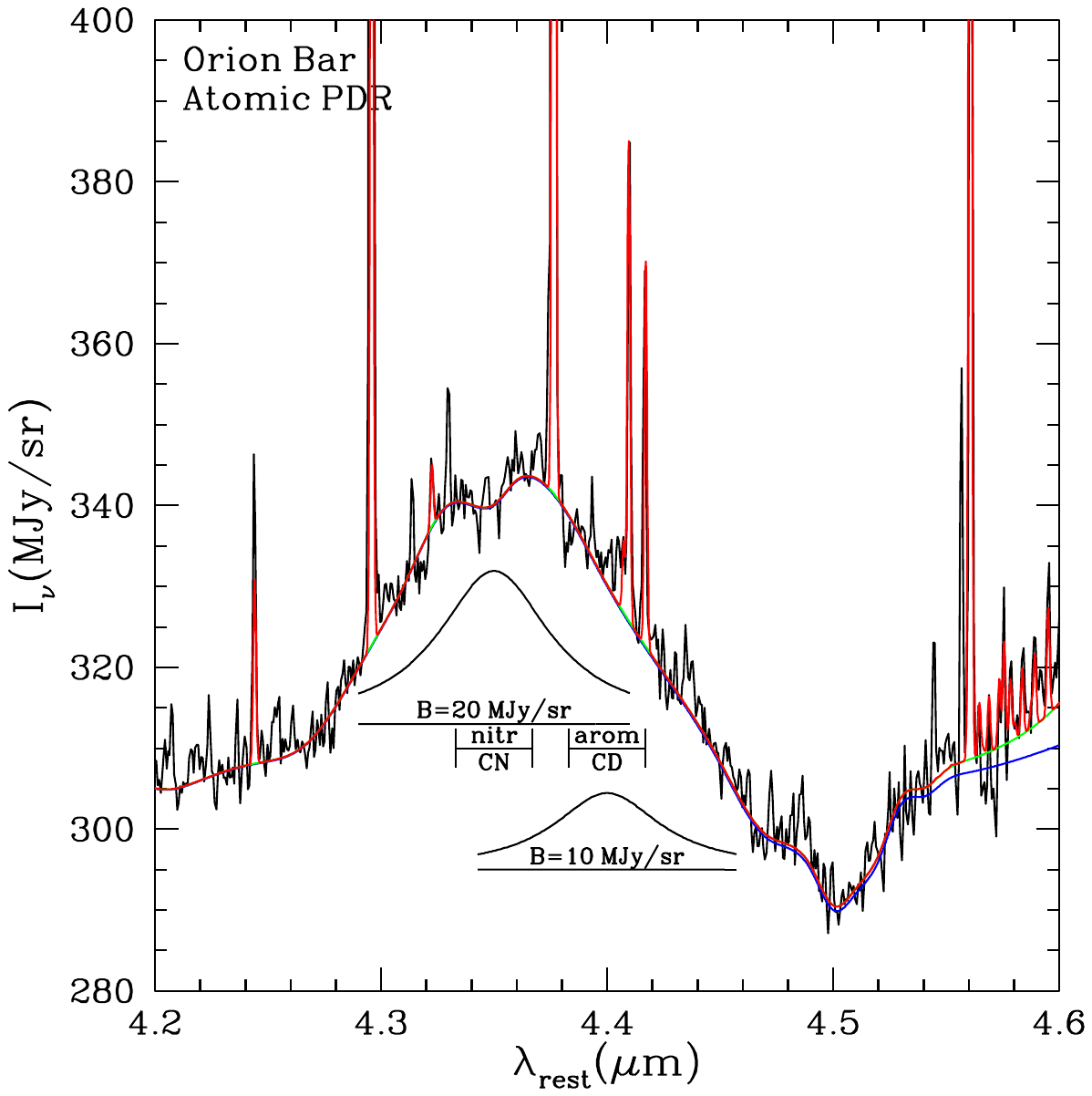}
\includegraphics[angle=0,width=\fwidth,
               clip=true,trim=0.5cm 5.cm 0.0cm 4.5cm]%
{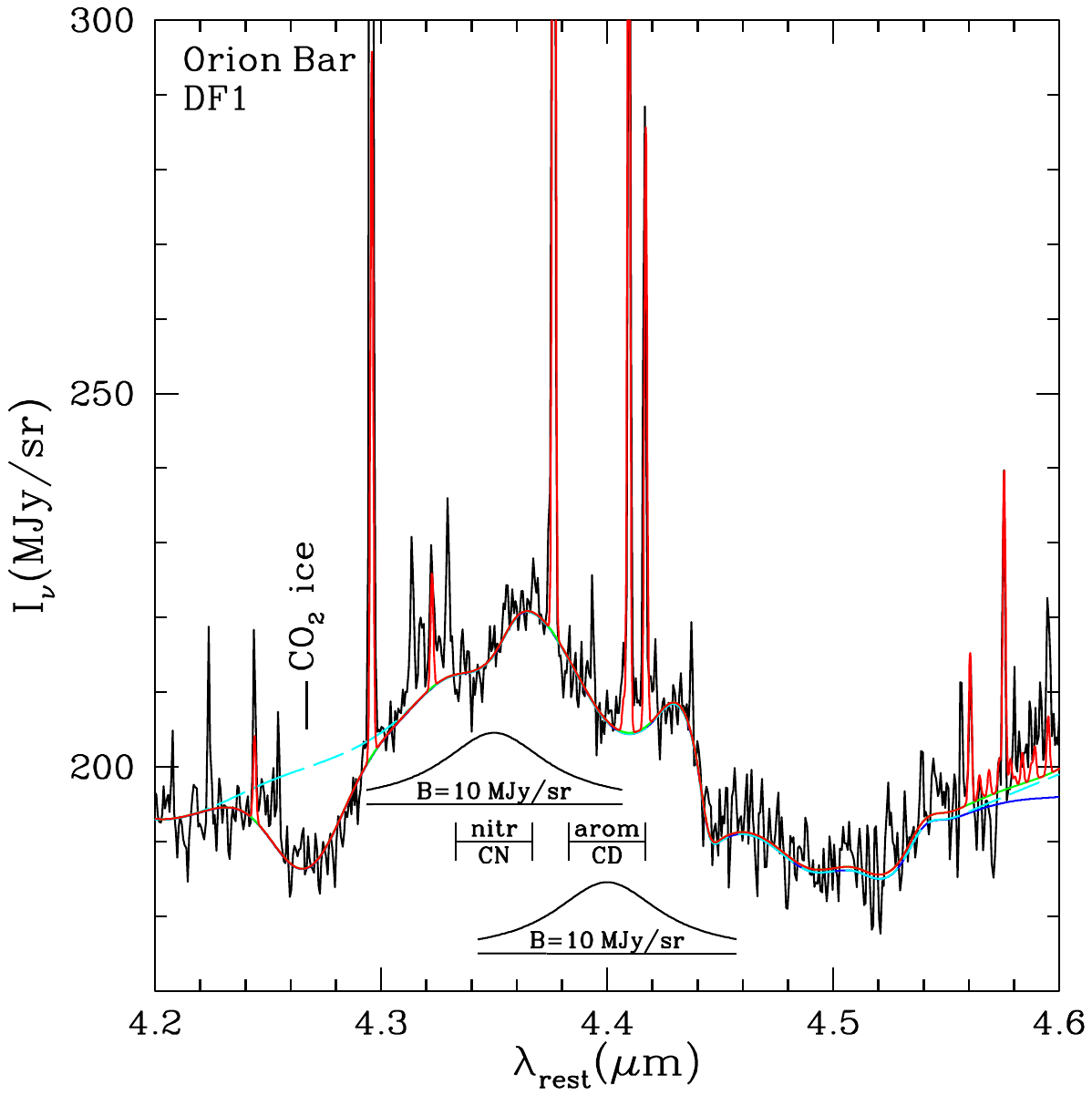}
\includegraphics[angle=0,width=\fwidth,
               clip=true,trim=0.5cm 5.cm 0.0cm 4.5cm]%
{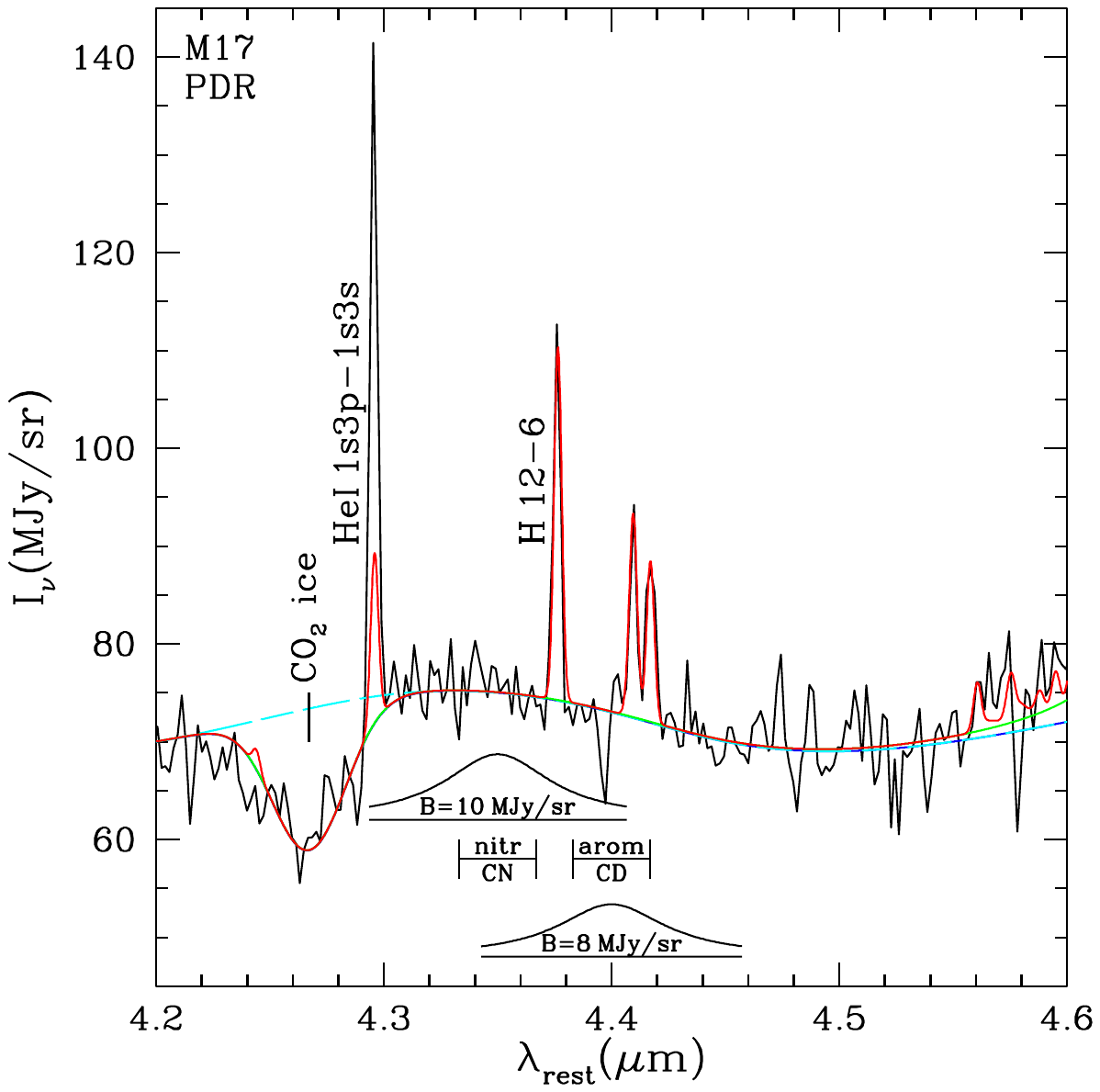}
\includegraphics[angle=0,width=\fwidth,
               clip=true,trim=0.5cm 5.cm 0.0cm 4.5cm]%
{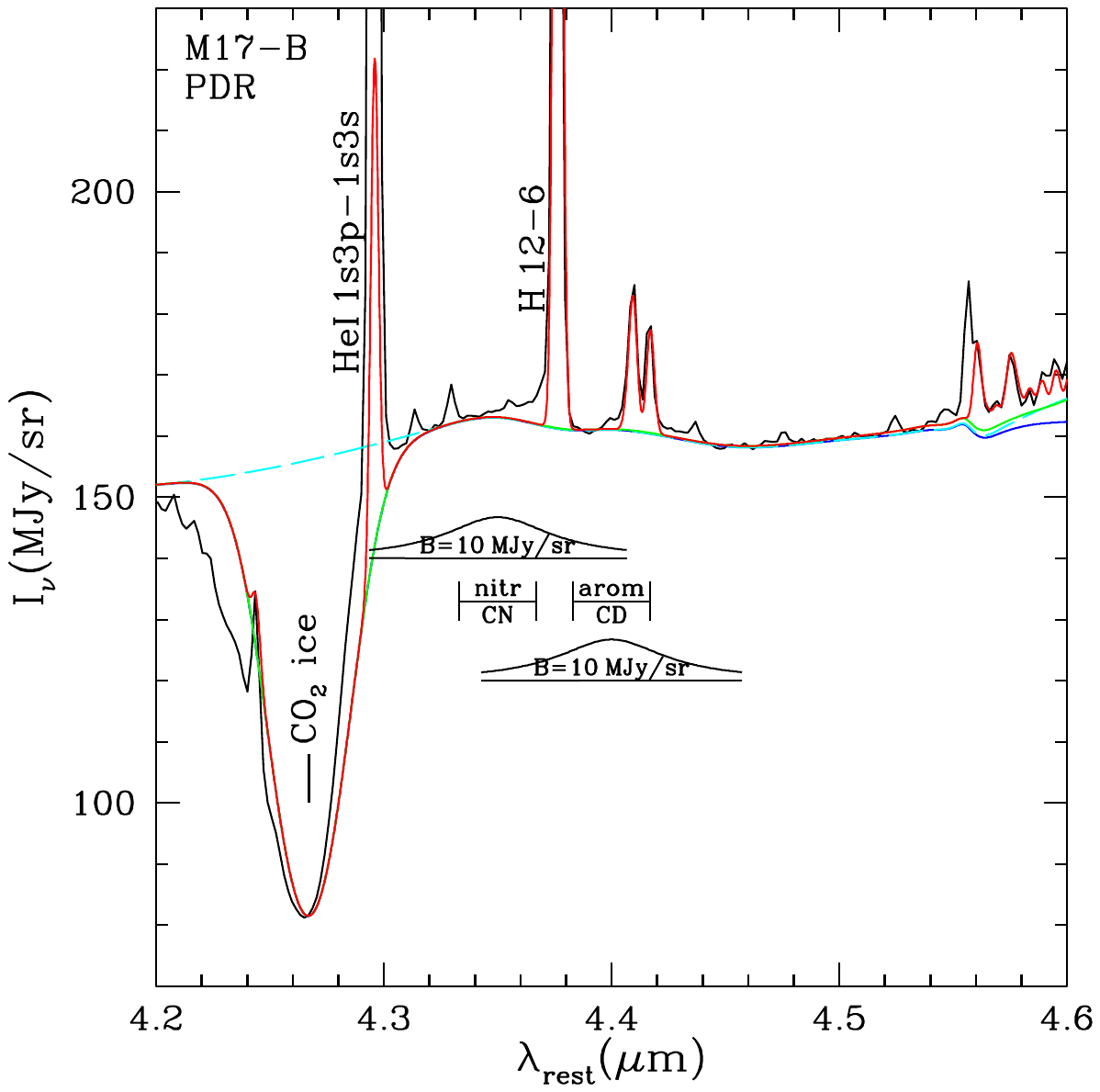}
\caption{\label{fig:MWaromCD}\footnotesize Same as Figure
  \ref{fig:m51aromCD}, but for two positions in the Orion Bar
  \citep{Peeters+Habart+Berne+etal_2024} and two positions in M17
  \citep{Boersma+Allamandola+Esposito+etal_2023}.}
\end{center}
\end{figure}

The aromatic C--D stretch is expected to be located at
$\lambda_\aromCD\approx4.40\micron$ \citep[][V.~J.~Esposito 2024,
  private communication]{Boersma+Allamandola+Esposito+etal_2023}.  As
discussed in Section \ref{sec:CH Stretch}, the $3.29\micron$ aromatic
C--H stretch can be approximated by a Drude profile (Equation
(\ref{eq:drude})) with $\gamma_\aromCH=0.013$.  The aromatic C--D
stretch resulting from substitution of D for H might be expected to
have similar dimensionless width $\gamma_\aromCD\approx 0.013$
($\FWHM\approx0.057\micron$).

An emission feature near $4.40\micron$ was tentatively detected in ISO
spectra of the Orion Bar
\citep{Peeters+Allamandola+Bauschlicher+etal_2004}, and appeared to be
present in $R\approx100$ AKARI spectra of a massive young stellar
object \citep{Onaka+Sakon+Shimonishi_2022}, with
$\FWHM\approx0.045\micron$.  However,
\citet{Boersma+Allamandola+Esposito+etal_2023} found no evidence for
an emission feature near $4.40\micron$ in JWST spectra of a number of
PDRs, and JWST spectra of the Orion Bar do not show a feature at
$4.40\micron$, instead showing a broad feature near $4.35\micron$ with
$\FWHM\approx0.1\micron$, which was suggested to be a combination of
the nitrile C-N stretch and the aromatic C--D stretch
\citep{Allamandola+Boersma+Lee+etal_2021,
  Peeters+Habart+Berne+etal_2024,Esposito+Fortenberry+Boersma+etal_2024a}.

Figure \ref{fig:m51aromCD} shows $4.2-4.6\micron$ spectra for the four
regions in M51.  There is no evidence
of an emission feature centered near the expected wavelength
$4.40\micron$ in any of the M51 spectra.  Each panel of Figure
\ref{fig:m51aromCD} includes a sample profile with
$\lambda_\aromCD=4.40\micron$, $\gamma_\aromCD = 0.013$, and an
amplitude $B_\aromCD^{\rm up.lim.}$ that would have made the feature
discernible in the observed spectrum (unless the underlying continuum
conspired to have a local minimum with similar central wavelength and
width).  We take the amplitude of these illustrative features to be an
upper limit on the power,
\beq
F_\aromCD^{\rm corr} = \frac{\pi c}{2}
\frac{\gamma_\aromCD }{\lambda_\aromCD}\times B_\aromCD
~~,
\eeq
in the aromatic C--D stretch feature.  Upper limits are given in Table
\ref{tab:params}.

The absence of a $4.40\micron$ feature due to aromatic C--D leads to
an upper limit on D/H in the aromatic material:
\beq
\left(\frac{\Da}{\Ha}\right)_\arom < 
K \times \frac{[F_\aromCD^{\rm\,corr}]_{\rm up.lim.}}{F_\aromCH^{\rm\,corr}}
=
K \times \frac{F_\aromCH^{\rm clip}}{F_\aromCH} \times
\frac{[F_\aromCD^{\rm\,corr}]_{\rm up.lim.}}{F_\aromCH^{\rm\,clip,corr}}
~~~.
\eeq
where we may use the same $K$ as defined in Equation (\ref{eq:K}).
Taking $K\approx1.2$ and estimating
$F_\aromCH^{\rm\,clip}\approx0.71F_\aromCH$ (see Figure
\ref{fig:33+34}), the upper limits on $F_\aromCD^{\rm\,corr}$ in Table
1 yield upper limits $(\Da/\Ha)_\arom< 0.006$ in SW-Strip,1, and
$(\Da/\Ha)_\arom < 0.007$ in NE-Strip (see Table \ref{tab:D/H}).
D-enrichment of the aliphatic material exceeds that of the aromatic
material by at least a factor 20 in these regions.

Similarly, Figure \ref{fig:MWaromCD} shows $4.2-4.6\micron$ spectra
for the Orion Bar and M17, with upper limits for the strength of any
$4.40\micron$ feature.  Limits on $(\Da/\Ha)_\arom$ in the Orion Bar
and M17 are given in Table \ref{tab:D/H}.

\subsection{Upper Limits on the Nitrile CN Stretching Mode}

The Orion Bar spectra appear to have a broad emission feature centered
near $\sim$$4.35\micron$, tentatively identified by
\citet{Peeters+Habart+Berne+etal_2024} as the CN stretch in nitrile
(--C$\equiv$N) groups.  However,
\citet{Boersma+Allamandola+Esposito+etal_2023} studied seven Galactic
regions, including the M17 PDR, and noted that there was little
evidence of a nitrile emission feature near the expected wavelength of
$\sim$$4.35\micron$.  Similarly, our M51 spectra (see Figure
\ref{fig:m51aromCD}) show no evidence of an emission feature near
$4.35\micron$.

If we assume the FWHM of the $4.35\micron$ nitrile CN stretch to be
similar to the $0.06\micron$ FWHM expected for the aromatic C--D
stretch (see the sample profiles in Figures \ref{fig:m51aromCD} and
\ref{fig:MWaromCD}), the upper bound on the power in the nitrile CN
stretch is similar to the upper bound estimated for the aromatic C--D
stretch.

\subsection{Upper Limits on a $4.75\micron$ Feature}

\citet{Doney+Candian+Mori+etal_2016} reported a feature at
$4.75\micron$, with $\FWHM=0.047\micron$ in Akari spectra of 6
Galactic \ion{H}{2} regions, which they attributed to the symmetric
aliphatic CD stretching mode.  We do not see evidence of this feature
in our spectra of \ion{H}{2} regions in M51.  Figure
\ref{fig:m51aliphCD} shows, for each spectrum, a $4.75\micron$ feature
that would have been detectable, and which we take as an upper limit
on the feature strength.  We find $F_{4.75\mu{\rm m}}^{\rm
  corr}/F_\aromCH^{\rm corr} < 0.004$ for M51 NE-Strip\,\ion{H}{2},
and $<0.011$ for all four regions in M51, below the values found by
\citet{Doney+Candian+Mori+etal_2016}, which ranged from $0.011$ for
M8, to $0.12$ for M17b.  

The Orion Bar spectra are more complex (see Figure
\ref{fig:orion+m17aliphCD}).  \citet{Peeters+Habart+Berne+etal_2024}
reported detection of the feature in Orion, but in our reanalysis of
their data we only claim upper limits (see Table \ref{tab:compare}).
We estimate $F_{4.75\mu{\rm m}}/F_\aromCH < 0.003$ for the atomic PDR,
and $<0.004$ for DF1.

\subsection{Model Limitations}

At the $D=7.5\Mpc$ distance of M51
\citep{Csornyei+Anderson+Vogl+etal_2023}, our 1\farcs5 diameter
extraction corresponds to 55 pc.  With the observed emission
originating from a number of subregions within the beam, characterized
by different foreground material, the assumption of a uniform screen
of dust is unrealistic.  However, differential extinction over the
$4-5\micron$ spectral range considered here is small, so that
inaccuracies in modeling the dust extinction are not expected to be
important for studying the $4.65\micron$ emission feature.  The
presence of the XCN ice absorption feature on some sightlines (most
prominently in the M51 E-Strip spectrum) does, however, introduce
uncertainty.

Uncertainties also arise from the treatment of CO absorption.  It is
unrealistic to assume a single column density $N_{{\rm CO},v=0}$
between us and the emitting $v=1$ CO, with the absorbing CO having a
Gaussian velocity distribution with $b=3\kms$, but with the $v=1$ CO
sufficiently shifted in velocity so that emission is not absorbed by
the $v=0$ CO.  Failure of the model to accurately reproduce the
weak CO emission or absorption features is unsurprising.

\subsection{Sequestration of Deuterium from the Gas Phase}

\citet{Friedman+Chayer+Jenkins+etal_2023} obtained accurate
measurements of $(\Da/\Ha)_\gas$ on 16 Galactic sightlines, with
median $(\Da/\Ha)_\gas \approx 17\ppm$, corresponding to
$\Da_\dust/\Ha_{\rm total}\approx6\ppm$ sequestered in the dust.  Two
sightlines had $(\Da/\Ha)_\gas < 9\ppm$, implying $\Da_\dust/\Ha_{\rm
  total} > 14\ppm$.

In the diffuse interstellar medium, the carbon in dust grains amounts
to $\Ca_\dust/\Ha_{\rm total} \approx 126\pm56\ppm$
\citep{Jenkins_2009, Hensley+Draine_2021}.  The degree of
hydrogenation is uncertain but must be appreciable, given observation
of the $3.4\micron$ aliphatic C--H stretch in absorption on diffuse ISM
sightlines.

%
%

Let $f_{\aliph\Ca}$ be the fraction of the C atoms that are in
aliphatic (--CH$_2$--) form, and let 
\beq
\phi_\aliph \equiv \left(\frac{\Da}{\Ha+\Da}\right)_\aliph
\eeq
be the deuterated fraction for the aliphatic material.  For the four
regions observed in M51 we estimate $(\Da/\Ha)_\aliph=0.17\pm0.02$,
i.e., $\phi_\aliph\approx 0.15\pm0.02$, for the nanoparticles
contributing to the emission at $\lambda < 5\micron$.

With two hydrogens per aliphatic C, the total D sequestered in dust is
\beqa
\frac{\Da_\dust}{\Ha_{\rm total}} &~=~& 
2 f_{\aliph\Ca} \, \langle\phi_\aliph\rangle \times
\frac{\Ca_\dust}{\Ha_{\rm total}}
\\
&=& 14\ppm 
\left(\frac{f_{\aliph\Ca}}{0.2}\right)
\left(\frac{\langle\phi_\aliph\rangle}{0.3}\right)
\left(\frac{\Ca_\dust/\Ha_{\rm total}}{120\ppm}\right)
~~~,
\eeqa
where $\langle\phi_\aliph\rangle$ is averaged over \emph{all} of the
aliphatic material, including that in the larger grains.  The
aliphatic fraction $f_{\aliph\Ca}$ is uncertain, but the strength of
the $3.4\micron$ absorption observed in the diffuse ISM
\citep{Chiar+Tielens+Adamson+Ricca_2013,Hensley+Draine_2020} requires
$f_{\aliph\Ca}\approx 0.3 (2.5\times10^{-18}\cm/A_{3.4\mu{\rm m}})$,
with band strengths $A_{3.4\mu{\rm m}}$ in the range
$(0.5-5)\times10^{-18}\cm$ estimated for various cations and neutrals
\citep[see, e.g.,][]{Yang+Li+Glaser+Zhong_2017}.  For
$f_{\aliph\Ca}\approx 0.2$ and $\langle\phi_\aliph\rangle > 0.15$, the
aliphatic hydrocarbons in the interstellar dust population can account
for the typical levels of ``missing D'' in the nearby interstellar
medium, while $\langle\phi_\aliph\rangle = 0.3$ would account for the
highest observed levels of ``missing D''.

%

\subsection{Detectability of Aliphatic C-D in Absorption}

The $4.65\micron$ feature may be detectable in absorption on diffuse
ISM sightlines with large $A_V$.  We estimate
\beq
\left[\int d\lambda^{-1}\Delta\tau \right]_{4.65} =
\frac{\langle\phi_\aliph\rangle}{1-\langle\phi_\aliph\rangle}
\frac{A_\CD}{A_\CH}
\left[\int d\lambda^{-1}\Delta\tau\right]_{3.42}
~~~,
\eeq
where $A_\CD/A_\CH\approx1/1.75$
\citep{Bauschlicher+Langhoff+Sandford+Hudgins_1997}.  With $A_V\approx
10.2$\,mag \citep{Torres-Dodgen+Tapia+Carroll_1991} and
$[\int d\lambda^{-1}\Delta\tau]_{3.42} = 3.90\cm^{-1}{\rm mag}$
\citep{Hensley+Draine_2020} toward Cyg OB-12, we estimate
\beq 
\left[\int d\lambda^{-1} \Delta\tau\right]_{4.65} 
\approx 0.22\cm^{-1} 
\frac{\langle\phi_\aliph\rangle}{1-\langle\phi_\aliph\rangle}
\left(\frac{A_V}{\rm mag}\right)
~~~,
\eeq
and central optical depth
\beq
(\Delta\tau)_{4.65\mu{\rm m}} \approx 0.011
\frac{\langle\phi_\aliph\rangle}{1-\langle\phi_\aliph\rangle}
\left(\frac{0.0265\micron}{\FWHM}\right) \left(\frac{A_V}{\rm
  mag}\right)
~~~.
\eeq
If $\langle\phi_\aliph\rangle \approx 0.3$, JWST NIRSpec observations
on heavily reddened diffuse ISM sightlines may be able to detect the
4.65$\micron$ aliphatic CD stretch in absorption.

\bigskip\bigskip

\section{\label{sec:summary} Summary}

Our principal conclusions are as follows:
\begin{enumerate}

\item The spectra of four massive star-forming regions in M51 include
  an emission feature near $4.65\micron$ attributed to emission in the
  aliphatic C--D stretching mode in PAH-related hydrocarbon nanoparticles.

\item The observed aliphatic C--D emission can be approximated by a
  Drude profile (Equation (\ref{eq:drude})) with
  $\lambda_\aliphCD\approx4.647\micron$, and
  $\FWHM\approx0.0265\micron$.

\item In the four regions of M51 studied here, the power in the aliphatic
  C--D emission feature ranges from 12\% to 18\% of the power in the
  ``clipped'' estimate of the $3.37-3.60\micron$ nonaromatic C--H
  emission -- 2 to 3 times the fraction found in the Orion Bar PDR.
  Evidently, the aliphatic material in PAH-related hydrocarbon
  nanoparticles in M51 is 2 to 3 times more deuterated than in the
  Orion Bar.  For reasonable assumptions concerning the temperature of
  the emitting nanoparticles, the aliphatic emitters in M51 have
  $\langle\Da/\Ha\rangle_\aliph \approx 0.17\pm0.02$ -- enrichment by
  a factor $\sim$$10^4$ relative to the overall D/H in the ISM.

\item Nondetection of deuterated aromatic emission near $4.40\micron$
  leads to upper limits on deuteration of the aromatic material.  We
  find $(\Da/\Ha)_\arom < 0.006$ in M51 SW-Strip,1, and $<0.003$ in
  the Orion Bar atomic PDR.  D enrichment of the aliphatic material
  exceeds that of the aromatic material by at least a factor 20.

\item None of the four positions in M51 show evidence of nitrile C-N
  emission at $4.35\micron$.  We estimate an upper limit
  $F_{4.35\micron}/F_\aromCH < 0.014$ for all four positions in M51.

\item None of the four positions in M51 show evidence of the
  $4.75\micron$ emission feature seen in AKARI spectra of Galactic
  \ion{H}{2} regions \citep{Doney+Candian+Mori+etal_2016}.  We
  estimate upper limits $F_{4.75\micron}/F_\aromCH < 0.011$ for all
  four positions in M51.

\item The ISM in M51 has O/H close to solar
  \citep{Croxall+Pogge+Berg+etal_2015}.  With similar degrees of
  ``astration'', D/H in M51 is presumed to be close to that in the
  solar neighborhood.  Differences in aliphatic deuteration among
  regions in M51, and relative to M17 and the Orion Bar, must result
  from different environmental conditions experienced by the
  hydrocarbon nanoparticles.

\item The aliphatic deuteration is found to anticorrelate with
  $\He^+/\Ha^+$ in the associated \ion{H}{2} (see Figure
  \ref{fig:cdr}b).  Harder and more intense FUV radiation from more
  massive stars may act to more rapidly de-deuterate the nanoparticles
  in PDRs near these stars.

\item The depletions of D observed in the diffuse ISM can be accounted
  for if all of the aliphatic material in dust is as deuterated as
  observed for the nanoparticles in M51,

\item The $4.65\micron$ aliphatic CD stretch may be detectable in
  absorption on sightlines to Galactic stars with large $A_V$.

\end{enumerate}

The JWST data presented in this Letter were obtained from the
MAST at the Space Telescope
Science Institute. The JWST observations analyzed can be accessed via
\dataset[doi:10.17909/k6kg-nc75]{https://doi.org/10.17909/k6kg-nc75} .
These observations are associated with JWST program 3435.

\begin{acknowledgements}

We thank G. DelZanna, V.~J.~Esposito and Aigen Li for helpful
discussions.  We especially thank P.~J.~Storey for kindly making
available extended tables of \ion{H}{1} emissivities.  We thank the
anonymous referee for a thoughtful and very helpful report that led to
improvements in the paper.

This work is based on observations made with the NASA/ESA/CSA James
Webb Space Telescope. The data were obtained from the Mikulski Archive
for Space Telescopes at the Space Telescope Science Institute, which
is operated by the Association of Universities for Research in
Astronomy, Inc., under NASA contract NAS 5-03127 for JWST. These
observations are associated with GO program \#3435.  The authors
acknowledge the PDRs4All Team for their publicly available data
products from the Early Release Science Program \#1288.  We also used
publicly available data for M17 from Cycle 1 GO Program 1591, PI
L.~Allamandola.
Support for program GO \#3435 was provided by NASA through a grant
from the Space Telescope Science Institute, which is operated by the
Association of Universities for Research in Astronomy, Inc., under
NASA contract NAS 5-03127.

This research has made use of NASA's Astrophysics Data System, as well
as ds9, a tool for data visualization supported by the Chandra X-ray
Science Center (CXC) and the High Energy Astrophysics Science Archive
Center (HEASARC) with support from the JWST Mission office at the
Space Telescope Science Institute for 3D visualization.

K.S., D.D., J.D.S., A.D.B., M.L.B., D.C., K.D.G., A.K.L., J.R.D., and
B.F.W. acknowledge support from grant JWST-GO-03435.
A.A. acknowledges support from the Swedish National Space Agency (SNSA)
through the grant 2021-00108.
R.S.K. acknowledges financial support from the European Research Council
via the ERC Synergy Grant ``ECOGAL'' (project ID 855130), from the
German Excellence Strategy via the Heidelberg Cluster of Excellence
(EXC 2181-390900948) ``STRUCTURES'', and from the German Ministry
for Economic Affairs and Climate Action in project ``MAINN'' (funding
ID 50OO2206). R.S.K. is grateful for computing resources provided by the
Ministry of Science, Research and the Arts (MWK) of the State of
Baden-W\"{u}rttemberg through bwHPC and the German Science Foundation
(DFG) through grants INST 35/1134-1 FUGG and 35/1597-1 FUGG, and also
for data storage at SDS@hd funded through grants INST 35/1314-1 FUGG
and INST 35/1503-1 FUGG. 

\end{acknowledgements}

\bibliography{/u/draine/work/libe/btdrefs}

\appendix

\section*{\label{app:CO abs} CO Lines in Absorption}

Let $N_{\rm CO}(v,J)$ be the column density of CO$(v,J)$.  The
line-center optical depth in an absorption line is
\beqa \tau_0 
&=&
\frac{1}{8\pi^{3/2}} \frac{2J_u+1}{2J_\ell+1} A_{u\ell}
\frac{\lambda_{u\ell}^3 N(v_\ell,J_\ell)}{b} \\ &=& 2.78
\frac{(2J_u+1)/3}{(2J_\ell+1)/1}
\left(\frac{A_{u\ell}}{11.5\s^{-1}}\right)
\left(\frac{\lambda_{u\ell}}{4.7576\micron}\right)^3
\left(\frac{N(v_\ell,J_\ell)}{10^{16}\cm^{-2}}\right)
\left(\frac{3\kms}{b}\right)
~~,
\eeqa
where $b$ is the usual Doppler-broadening parameter.  The correction
for stimulated emission is neglected because (1) the population in the
$v=1$ levels is small, and (2) the velocity distribution of the $v=1$
CO is likely to differ from that of the $v=0$ CO.  Because the cold CO
is likely at a different velocity than the hot $v=1$ CO, we do not
apply CO absorption to the CO emission lines.

The dimensionless equivalent width in the absorption line is taken to be
\citep[][Eq. 9.27]{Draine_2011a}
\beqa
W 
&\approx&
\sqrt{\pi} ~ \frac{b}{c}~ \frac{\tau_0}{1+\tau_0/(2\sqrt{2})} ~~~{\rm for}~~~\tau_0 \leq 1.25393
\\
  &\approx& \frac{2b}{c} \left[\ln\left(\frac{\tau_0}{\ln2}\right)\right]^{1/2}  ~~~~{\rm for}~~~\tau_0 > 1.25393
~~~.
\eeqa
The JWST spectra do not resolve the absorption line.  Let
$R\equiv\lambda/\FWHM_\lambda$ be the resolution of the spectrograph.
For $N({\rm CO})\ltsim 10^{17}\cm^{-2}$ and $b\approx 3\kms$, we have
$\tau_0\ltsim 30$ and $W\ltsim 4\times10^{-5}$.  If $W \ll 1/R$ (as
will be the case for JWST resolution $R\ltsim 3000$), the attenuation
due to the line is
\beq
\alpha(\lambda) = 
\frac{W}{\sqrt{2\pi}\,\sigma}
e^{-(\lambda-\lambda_{u\ell})^2/2\sigma_\lambda^2}
~~~,
\eeq
where $\sigma=1/(R\sqrt{8\ln2})$, 
and $\sigma_\lambda = \lambda/(R\sqrt{8\ln2})$.

\end{document}